\documentclass[10pt,superscriptaddress,twocolumn,amsmath,amssymb,aps,prb]{revtex4}
\usepackage{mathrsfs}
\usepackage{graphicx}% Include figure files
\usepackage{dcolumn}% Align table columns on decimal point
\usepackage{bm}% bold math
\newcommand{\be}{\begin{equation}}
\newcommand{\ee}{\end{equation}}
\newcommand{\bea}{\begin{eqnarray}}
\newcommand{\eea}{\end{eqnarray}}

\begin{document}

\title{ Magnetism and superconductivity in the layered hexagonal transition metal pnictides}
\author{Jinfeng Zeng}
\affiliation{Beijing National Laboratory for Condensed Matter Physics,
and Institute of Physics, Chinese Academy of Sciences, Beijing 100190, China}
\affiliation{University of Chinese Academy of Sciences, Beijing 100049, China}

\author{Shengshan Qin}
\affiliation{Beijing National Laboratory for Condensed Matter Physics,
and Institute of Physics, Chinese Academy of Sciences, Beijing 100190, China}
\affiliation{University of Chinese Academy of Sciences, Beijing 100049, China}

\author{Congcong Le}
\affiliation{Beijing National Laboratory for Condensed Matter Physics,
and Institute of Physics, Chinese Academy of Sciences, Beijing 100190, China}
\affiliation{Kavli Institute of Theoretical Sciences, University of Chinese Academy of Sciences,
Beijing, 100190, China}

\author{Jiangping Hu}\email{jphu@iphy.ac.cn}
\affiliation{Beijing National Laboratory for Condensed Matter Physics,
and Institute of Physics, Chinese Academy of Sciences, Beijing 100190, China}
\affiliation{Kavli Institute of Theoretical Sciences, University of Chinese Academy of Sciences,
Beijing, 100190, China}
\affiliation{Collaborative Innovation Center of Quantum Matter,
Beijing, China}

\date{\today}

\begin{abstract}
We investigate the electronic and magnetic structures of the 122 (AM$_2$B$_2$) hexagonal transition-metal pnictides with A=(Sr, Ca), M=(Cr, Mn, Fe, Co, Ni) and B=(As, P, Sb). It is found that the family of materials share  critical similarities  with those  of tetragonal structures that  include the famous  iron-based high temperature superconductors. In both families, the  next nearest neighbor(NNN)  effective antiferromagnetic(AFM) exchange couplings reach  the maximum value in the iron-based materials.  While the NNN couplings in the latter are known to be responsible for the C-type AFM state and to result in the extended s-wave superconducting state upon doping,  they  cause the former to be extremely frustrated magnetic systems  and can  lead to an  time reversal symmetry broken $d+id$ superconducting state upon doping. The iron-based compounds with the hexagonal structure,  thus if synthesized,  can help us to determine the origin of high temperature superconductivity.
\end{abstract}


\maketitle
\section{Introduction}\label{sectioni}
The accidental discovery of iron-based superconductors in 2008 \cite{Fe}  was a great surprise to the entire high temperature(high T$_c$)  superconductivity research community. Since then, the iron-based superconductors have been one of the most active research fields in condensed matter physics. It was wildly cheered that the study of the materials might eventually lead us to understand the superconducting mechanism of unconventional high temperature superconductors. Nevertheless, even if many rich physics in these materials have been discovered and well understood,  the superconductivity mechanism remains a controversial subject.

Theoretically, to understand high T$_c$ superconductivity, different electronic properties or phenomena have been selected and emphasized in different approaches  and models\cite{Hirschfeld2011}. The essential difficulty in solving high T$_c$ mechanism lies on how to identify  indispensable features that are directly tied to high T$_c$ superconductivity among complex electronic structures and physical phenomena. In principle, a successful identification should also lead us to predict new families of high T$_c$ superconductors.

If we assume that there should be one unified superconducting mechanism for unconventional high temperature superconductors including cuprates\cite{Cu}, it is possible to make the identification by comparing different classes of materials. Considering the 122 family of iron-based superconductors, the iron atoms can be fully replaced by other transition metal atoms, such as Cr, Mn, Co, Ni and Cu. These compounds have the same lattice structures as the 122 iron-based superconductors. Accumulative experimental evidences\cite{Sefat2009-Cr, Kasinathan2009-Mn, Pandey2011-Mn, Pandey2012prl-Mn, Ahilan2014-Co, Anand2014-Co, Ronning2008-Ni, ZhangPan2017-Ni,  Saparov2012-Cu, Anand2012-Cu} suggest that they do not exhibit high T$_c$ superconductivity.  These facts lead us to ask a profound question: \textit{why is iron so special  for high T$_c$ superconductivity?}

If we compare all these similar materials, only the iron-based materials exhibit the C-type AFM order\cite{DaiPengchengRMP2015} and the superconductivity emerges when the magnetic order is suppressed. The C-type AFM state indicates  the presence of strong effective  AFM exchange couplings between two NNN iron atoms. The NNN AFM couplings are known to stem from the superexchange mechanism through the couplings between the t$_{2g}$ iron d-orbitals and the anion p-orbitals.  In many previous theoretical studies\cite{Seo2008,Fang2011, Hu-Yuan}, these AFM interactions are shown to generate strong superconducting pairings with  extended s-wave symmetry. Recently we have also  pointed out that the $d^6$ configuration at Fe$^{2+}$ is an unique configuration to  isolate the $t_{2g}$ orbitals near Fermi energy\cite{jpHuprx, jpHusciencebult}. Thus, the special electronic structure in which  the t$_{2g}$  orbitals  are isolated near Fermi energy to generate the maximum superexchange AFM interactions in the vicinity of the $d^6$ configuration at Fe$^{2+}$ is suggested to be  the key  the high T$_c$ mechanism. This speciality is also satisfied in cuprates\cite{Cu} in which the single d$_{x^2-y^2}$ e$_g$ orbital of Cu$^{2+}$ which is responsible for the superexchange interactions is isolated near Fermi energy.

In this paper, we argue that the hexagonal 122 transition-metal pnictides, AM$_2$B$_2$ with A=(Sr, Ca), M=(Cr, Mn, Fe, Co, Ni) and B=(As, P, Sb), which have the trigonal CaAl$_2$Si$_2$-type structure, can be a new family of materials to test the above identification of the superconducting mechanism.  We compare the magnetic properties between the tetragonal and hexagonal 122 families of pnictides obtained from density functional theory(DFT) calculations and find that the overall trend of magnetism in the hexagonal structure as the change of transition metal atoms  is very similar to  the one in the tetragonal structure.  Their similarities include: (1) the NNN AFM exchange interactions reach maximum in Fe-based materials; (2) the nearest neighbor(NN) AFM interactions, which are mainly attributed to direct magnetic exchange mechanism, are very strong in Cr/Mn based materials; (3) in Co/Ni/Cu based materials, magnetic interactions are very weak or negligible. Because of the strong NNN AFM interactions, the iron-based hexagonal materials are extremely frustrated magnetic systems.  Their electronic structure near Fermi energy is mainly attributed to $t_{2g}$ orbitals that form two quasi two-dimensional electron pockets. Upon doping, the superconductivity with d+id spin singlet pairing symmetry can be developed.

The paper is organized as following. In Section~\ref{S2}, we present the calculation results  from density functional theory (DFT) and study the crystal structure of hexagonal 122 transition-metal pnictides AM$_2$B$_2$(A=Sr, Ca; M=Cr, Mn, Fe, Co, Ni; B=As, P, Sb).  In Section~\ref{S3}, we  review and summarize the magnetic properties of the tetragonal family. In Sec.~\ref{S4}, we investigate the effective magnetic exchange interactions in the hexagonal family. In Sec.~\ref{S5}, we investigate the  pressure effect on the magnetism of the hexagonal CaFe$_2$As$_2$. In Sec.~\ref{S6}, we discuss the electronic structures of  the hexagonal CaFe$_2$As$_2$ and analyze the  possible superconducting  state.  Finally, Sec.~\ref{S8}, we give a summary and provide the main conclusions of our paper.

\section{ Crystal structures and calculation methods} \label{S2}
\begin{figure}[tb]
\centerline{\includegraphics[width=0.5\textwidth]{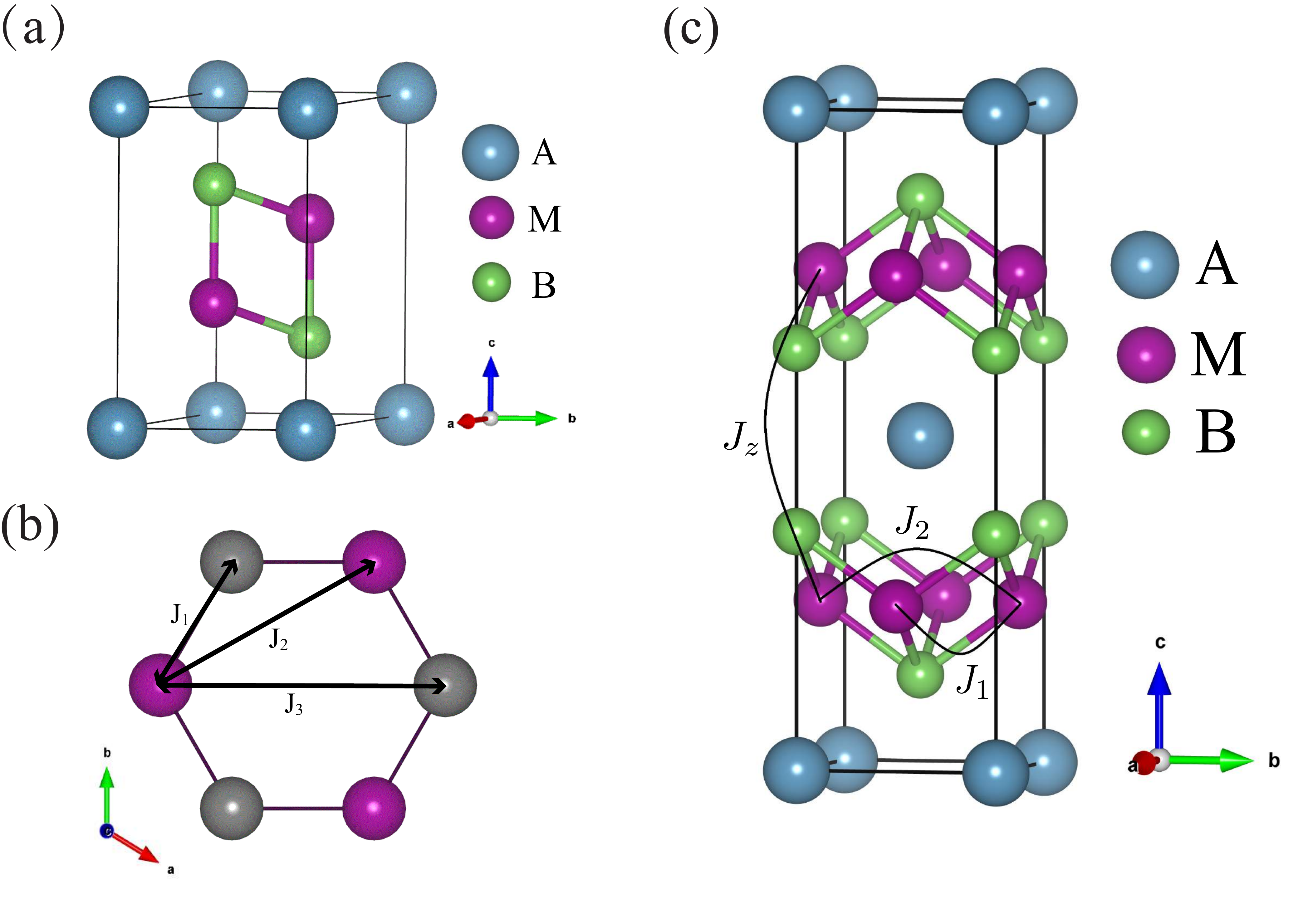}}
\caption{(color online). (a) The crystal structure of  the hexagonal AM$_2$B$_2$(A=Sr, Ca; M=Cr, Mn, Fe, Co, Ni; B=As, P, Sb), the trigonal CaAl$_2$Si$_2$ type structure(space group $P\bar{3}$m1, No. 164); (b) the corrugated honeycomb lattice formed only by the M atoms in the $ab$ plane. Exchange interactions between NN $J_1$, NNN $J_2$, and third NN $J_3$ are indicated; (c) the crystal structure of AM$_2$B$_2$ (A=Ba; M=Cr, Mn, Fe, Co, Ni; B=As, P) with  the body-centered-tetragonal structure(space group $I4/mmm$, No. 139). The in-plane NN magnetic  exchange interaction $J_1$,  the in-plane NNN exchange interaction $J_2$ and the out-of-plane NN along $c$ axis exchange interaction $J_z$ are indicated.}
\label{CrystalStructure-all}
\end{figure}

The 122 tetragonal and hexagonal crystal structures are shown in Fig. \ref{CrystalStructure-all}. The 122 iron-based superconductors, such as BaFe$_2$As$_2$, has a body-centered-tetragonal structure as shown in Fig. \ref{CrystalStructure-all}(c) with space group  $I4/mmm$.  Besides iron-based materials, materials with this crystal structure have been synthesized for Cr, Mn, Co, Ni and Cu  as well. The hexagonal 122 structure is shown in Fig.\ref{CrystalStructure-all}(a).  This structure is referred as the trigonal CaAl$_2$Si$_2$-type structure with space group $P\bar{3}$m1. (Sr, Ca)Mn$_2$As$_2$  with this structure have been synthesized. Here we use the DFT calculations to  systematically investigate these families of materials.  The 122 tetragonal family has been intensively investigated\cite{}.  The magnetic states have been correctly obtained by the DFT calculations\cite{}. Although the hexagonal 122 structure has also been investigated, there is no systematic DFT results.

Our DFT calculations employ the projector augmented wave (PAW) method encoded in Vienna \emph{ab initio} simulation package(VASP)\cite{Kresse1993,Kresse1996,Kresse1996B}, and generalized-gradient approximation (GGA)\cite{Perdew1996} for the exchange correlation functional is used.
We relax the lattice constants and internal atomic positions for hexagonal family, where the plane wave cutoff energy is 600 eV and these $k$ points are $13\times13\times7$. Forces are minimized to less than 0.01 eV/\AA~ in the structural relaxation. Throughout this work, the cutoff energy of 500 eV  for tetragonal family and 450 eV for hexagonal family are taken for expanding the wave functions into plane-wave basis. The number of these $k$ points are $7\times7\times3$ for tetragonal family and $7\times11\times7$ for hexagonal family in the calculation of the magnetic structures. The GGA plus on-site repulsion $U$ method (GGA$+U$) in the formulation of Dudarev {\it et al}.\cite{Dudarev1998} is employed to describe the electron correlation effect associated.

\begin{table}[tb]
\addtolength{\tabcolsep}{13pt}
\caption{The experimental crystal structure parameters for BaM$_2$B$_2$(M=Cr, Mn, Fe, Co, Ni; B=P, As) with the body-centered-tetragonal structure (space group $I4/mmm$). }
\begin{tabular}{ccccc}
\hline
\hline
                                           & $a$(\AA)       & $c$(\AA)    & $Z_{As}$  \\
\hline
BaCr$_2$As$_2$\cite{BaCr2As2andBaCo2As2}    & 3.963         & 13.600    & 0.3572  \\
%\hline
BaMn$_2$As$_2$\cite{BaMn2As2}              & 4.154        & 13.415    & 0.3613  \\
%\hline
BaFe$_2$As$_2$\cite{BaFe2As2}             &  3.963       & 13.017  & 0.3545  \\
%\hline
BaCo$_2$As$_2$\cite{BaCr2As2andBaCo2As2}  &  3.958       & 12.670  & 0.3509  \\
%\hline
BaNi$_2$As$_2$\cite{BaNi2As2}             &    4.112    &  11.540 &  0.3476 \\
%\hline
BaFe$_2$P$_2$\cite{BaFe2P2}               &     3.840    & 12.442  &  0.3456\\
\hline
\hline
\end{tabular}
\label{tab:cs-tetragonal}
\end{table}

In  the 122 tetragonal family,  the A site has very limited effect on electronic physics. So in our calculations and analysis, we set A site to Ba atom.   We adopt  the experimental lattice constants in our calculations as all those materials have been synthesized are shown in table \ref{tab:cs-tetragonal}. For the hexagonal family, CaMn$_2$As$_2$, CaMn$_2$P$_2$, CaMn$_2$Sb$_2$, SrMn$_2$P$_2$ and SrMn$_2$As$_2$ have been synthesized\cite{CaMn2As2, CaMn2P2, SrMn2P2, CaMn2Sb2}. Similar to the tetragonal family structure, the A site atoms have little affect on the electronic and magnetic structure near the fermi level. In the following calculations and analysis, we set A site to Ca atom. The trigonal CaAl$_2$Si$_2$-type structure is shown in Fig. \ref{CrystalStructure-all}(a). The M site sublattice forms the corrugated honeycomb lattice, which is shown in Fig. \ref{CrystalStructure-all}(b). The purple and gray M atoms belong to the up and low trigonal layers respectively.  The lattice constants are listed in table \ref{tab:cs-exp-relax}. CaMn$_2$As$_2$  has been reported with antiferromagnetic(AF) order\cite{Sangeetha2016}. However, the other materials in CaM$_2$B$_2$(M=Cr, Fe, Co, Ni) have not been synthesized. In order to study their magnetic structures, we  relax the lattice constants and internal atomic positions with the GGA$+U$ method in the AFM ordered states. By relaxing the lattice constants and internal atomic positions for CaMn$_2$As$_2$ and CaMn$_2$P$_2$ with $U_{eff}$=(0, 0.5, 1, 1.5, 2, 2.5) eV, we find that when $U_{eff}$=1.5 eV, the optimized lattice constants are the most closed to the experimental data for both CaMn$_2$As$_2$ and CaMn$_2$P$_2$, which are given in Table \ref{tab:cs-exp-relax}.  Therefore, we adopt the value $U_{eff}$=1.5 eV to relax the other materials CaM$_2$As$_2$(CaM$_2$P$_2$)(M=Cr, Fe, Co, Ni) with the experimental lattice constants of CaMn$_2$As$_2$(CaMn$_2$P$_2$) as the input parameters. The results of the optimized structural parameters are listed in Table \ref{tab:cs-hexagonal}.
In the calculations of the magnetic states of the trigonal CaAl$_2$Si$_2$-type structure CaM$_2$B$_2$, we double the primitive cell as unit cell, which is indicated by the red frame in Fig. \ref{magnetic-states}.

\begin{table}[tb]
\addtolength{\tabcolsep}{3pt}
\caption{Experimental and optimized structural parameters of CaMn2B2(B=As, P) by using GGA$+U$($U_{eff}=1.5$ eV) in the AFM ordered state.}
\begin{tabular}{ccccccc}
\hline
\hline
%   & Lattice parameters & &  &  Atomic coordinates & \\
%\hline
   & $a$(\AA)  & $c$(\AA) & $z_{X}$ & $z_{Y}$ \\
\hline
CaMn$_2$As$_2$(experiment)\cite{CaMn2As2}     &4.230  & 7.030   & 0.6237 & 0.2557  \\
%\hline
CaMn$_2$As$_2$(relax)   &4.258  & 7.002   & 0.6203 & 0.2574  \\
\hline
CaMn$_2$P$_2$(experiment)\cite{CaMn2P2}       &4.096  & 6.848   & 0.6246 & 0.2612  \\
%\hline
CaMn$_2$P$_2$(relax)    &4.109  & 6.775   & 0.6213 & 0.2644  \\
\hline
\hline
\end{tabular}
\label{tab:cs-exp-relax}
\end{table}

\begin{table}[tb]
\addtolength{\tabcolsep}{9pt}
\caption{The optimized crystal structure parameters for CaM$_2$B$_2$(M=Cr, Mn, Fe, Co, Ni; B=P, As) with  the trigonal CaAl$_2$Si$_2$-type structure(space group $P\bar{3}$m1) by using GGA$+U$($U_{eff}=1.5$ eV) in the AFM ordered state.}
\begin{tabular}{ccccccc}
\hline
\hline
%   & Lattice parameters & &  &  Atomic coordinates & \\
%\hline
   & $a$(\AA  & $c$(\AA) & $z_{X}$ & $z_{Y}$ \\
\hline
CaCr$_2$As$_2$  &4.122  & 7.269   & 0.6180 & 0.2553  \\
%\hline
CaMn$_2$As$_2$  &4.258  & 7.002   & 0.6203 & 0.2574  \\
%\hline
CaFe$_2$As$_2$  &4.086  & 6.834   & 0.6237 & 0.2740  \\
%\hline
CaCo$_2$As$_2$  &3.891  & 6.761   & 0.6269 & 0.2894  \\
%\hline
CaNi$_2$As$_2$  &3.965  & 6.714   & 0.6299 & 0.2869  \\
\hline
CaCr$_2$P$_2$   &3.972  & 7.023   & 0.6203 & 0.2630  \\
%\hline
CaMn$_2$P$_2$   &4.109  & 6.775   & 0.6213 & 0.2644  \\
%\hline
CaFe$_2$P$_2$   &3.858  & 6.639   & 0.6236 & 0.2859  \\
%\hline
CaCo$_2$P$_2$   &3.715  & 6.602   & 0.6263 & 0.2973  \\
%\hline
CaNi$_2$P$_2$   &3.788  & 6.558   & 0.6271 & 0.2919  \\
\hline
\hline
\end{tabular}
\label{tab:cs-hexagonal}
\end{table}

\section{The Magnetism in the tetragonal 122 transition-metal pnictides}\label{S3}
\begin{figure}[tb]
\centerline{\includegraphics[width=0.5\textwidth]{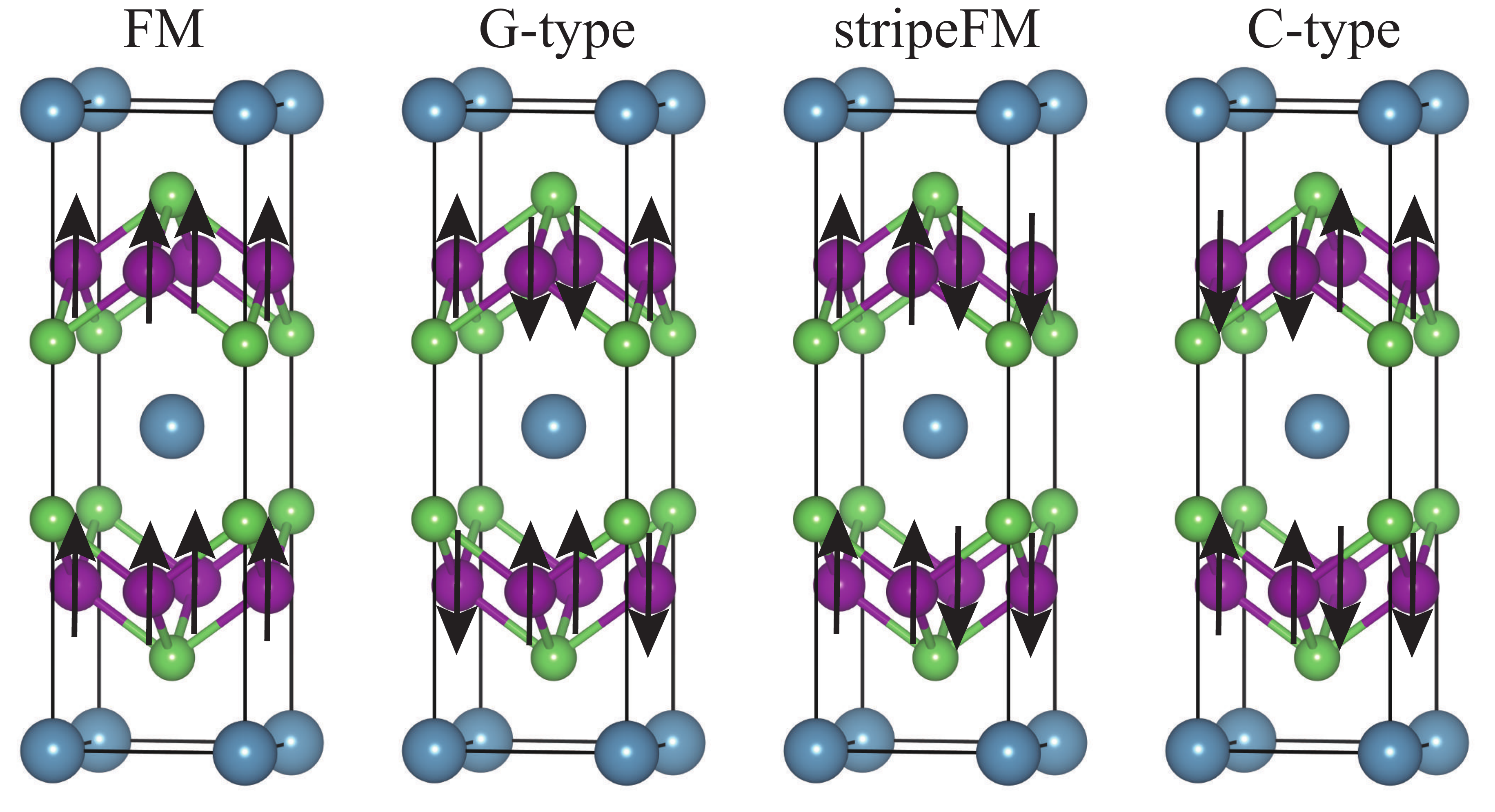}}
\caption{(color online) The sketch of  the four collinear magnetic states, including the FM states, the G-type AFM state, the stripeFM state and the C-type AFM state.}
\label{tetragonal-magnetic-states}
\end{figure}

In this section, we review and investigate the magnetic properties of the 122 tetragonal BaM$_2$B$_2$(M=Cr, Mn, Fe, Co, Ni; B=As, P). We consider four competing collinear magnetic states, the ferromagnetic(FM) state, the G-type AFM state and two stripe states. We distinguish two stripe states as the stripe FM  state with ferromagnetic alignment of adjacent spins along the c axis and the C-type AFM state with antiferromagnetic alignment. The above four magnetic states are shown in Fig. \ref{tetragonal-magnetic-states}. It has been known that the magnetism can be described by the effective $J_1-J_2-J_z$ Heisenberg model\cite{Fang2008, Fang2009, reviewDCJohnston, Holt2011, Stanek2011, Johnston2011}, which is given by
\begin{equation}
\begin{aligned}
\emph{H}=J_{1}\sum_{\langle ij \rangle}S_{i}\cdot S_{j} + J_{2}\sum_{\langle\langle ij \rangle\rangle}S_{i}\cdot S_{j}+J_{z}\sum_{\langle ij \rangle_c}S_{i}\cdot S_{j}
\end{aligned}
\label{hh}
\end{equation}
whereas $\langle ij \rangle$ , $\langle\langle ij \rangle\rangle$ and $\langle ij\rangle_c$ denote the summation over the in-plane NN, in-plane NNN and out-of-plane NN along $c$ axis, respectively. The exchange interaction parameters $J_1$, $J_2$ and $J_z$ are indicated in Fig. \ref{CrystalStructure-all}(c). $S_{i}$ is the spin operator for the $i$th site.
Throughout this paper, a positive $J$ corresponds to an antiferromagnetic interaction and a negative $J$ to a ferromagnetic interaction.
The classical energies of the above magnetic states are
\begin{eqnarray}
\label{tetragonal-energy-to-magnetic-state}
E_{FM} & = & NS^2(2J_{1} + 2J_{2} + J_{z}) + E_{0} \nonumber\\
E_{G-type}  & = & NS^2(-2J_{1} + 2J_{2} - J_{z})+ E_{0} \nonumber \\
E_{stripeFM} & = & NS^2(-2J_{2} + J_{z})+ E_{0}     \nonumber \\
E_{C-type} & = & NS^2(-2J_{2} - J_{z})+ E_{0}
\end{eqnarray}
where $E_0$ is the energy of nonmagnetic state.

We calculate the magnetic moments and the total energies of the above magnetic states with GGA$+U$.   If the calculated local magnetic moments at M sites are very close to each other, we can extract the magnetic exchange coupling constants by Eq. (\ref{tetragonal-energy-to-magnetic-state}). The results show that, in BaCr$_2$As$_2$ the average moments of the four magnetic states  vary $2.3-3.4 \mu_{B}$. The values increase when  $U_{eff}$ increases. The order moments do not vary significantly  in the above four magnetic states. The average moments vary $3.5-4.1\mu_{B}$ in  BaMn$_2$As$_2$ and $2.0-2.9\mu_{B}$ in BaFe$_2$As$_2$. For BaCo$_2$As$_2$, the moments are very small.  Finally, in the BaNi$_2$As$_2$, the moments are zero within the range of error $0.001\mu_{B}$ and there are no energy gain for magnetic states.  These results are consistent with previous calculations\cite{Singh2009, Filsinger2017, Johnston2011, AnJiming2009, Ma2010, XuGang2008, Zbiri2009, Yin2009, Boeri2010, Han2009, Sefat2009-Co}. For BaNi$_2$As$_2$, our results consistent with the angle-resolved photoemission spectroscopy experiment\cite{NiARPESNoMoment}, which show that collinear spin-density-wave magnetic ordering does not exist in BaNi$_2$As$_2$.

We can extract the magnetic exchange parameters by Eqs. (\ref{tetragonal-energy-to-magnetic-state}).   The results are summarized below.  For Co-based materials, as the magnetic moment is too small, it is not reliable to extract these exchange parameters. For Ni-based materials,  the calculated magnetic exchange parameters are zero. For other three materials, the results are consistent with previous calculations as well as experimental measurements.   As the energy gain in the magnetic state is proportional to $J<S>^2$ and the ordered magnetic moments are large in all three materials, we can assume the atoms are close to  the high spin states for simplicity.  In the above effective model, we take the spin values to be the high spin of the atoms.   For example, the Cr$^{2+}$ ion has four electron so that the spin S=2.  Similarly, we take S=$\frac{5}{2}$ in Mn$^{2+}$ ion, S=2 in Fe$^{2+}$ ion.

In the case of $U_{eff}$ = 0, our calculated $J_2$/$J_1$ value of BaCr$_2$As$_2$ is $-0.74$. It is very close to $-0.85$, which is give in the previous work \cite{Singh2009}. And our results show that $J_1 > 0 $ and $J_2 < 0 $ in all $U_{eff}$ values, which suggests that the BaCr$_2$As$_2$ has G-type AFM order. This is consistent with the theoretical calculations\cite{Singh2009, Richard2017} and powder neutron diffraction experiment\cite{Filsinger2017}. In BaMn$_2$As$_2$, our calculated exchange parameters are $J_1$ = 15.30 meV, $J_2$ = 2.33 meV and $J_z$ = 1.08 meV with high spin values for $U_{eff}$ = 0, which is very close to the values given in the previous calculation\cite{Johnston2011}.  We also obtain that $J_1$/$J_2$ $ <  \frac{1}{2}$, which suggests that the BaMn$_2$As$_2$ has G-type AFM order. And in our calculation, the G-type AFM order has the lowest energy among the four magnetic order given above. Our results are consistent with the theoretical calculations\cite{Johnston2011} and neutron diffraction measurements\cite{Singh2009-Mn}. The neutron diffraction experiment also shows that the ordered moment is $3.88(4)\mu_{B}$\cite{Singh2009-Mn}. Our results give that the ordered moment is $3.856\mu_{B}$ in the G-type state with $U_{eff}$ = 1.5eV. In the case of $U_{eff}$ = 0, our calculated exchange parameters are $J_1 S^2$ = 31.39 meV and  $J_2 S^2$ = 33.51 meV for BaFe$_2$As$_2$ , which is similar to the values $J_1  S^2$ = 25.5 meV and $J_2  S^2$ = 33.8 meV given in the previous calculations\cite{Ma2010}. $J_1$/$J_2$ $> \frac{1}{2}$  in all $U_{eff}$ values, which suggests that BaFe$_2$As$_2$ has a C-type AFM order state. These results are consistent with the neutron diffraction experiment measurements\cite{Huang2008}.

\begin{figure}[tb]
\centerline{\includegraphics[width=0.5\textwidth]{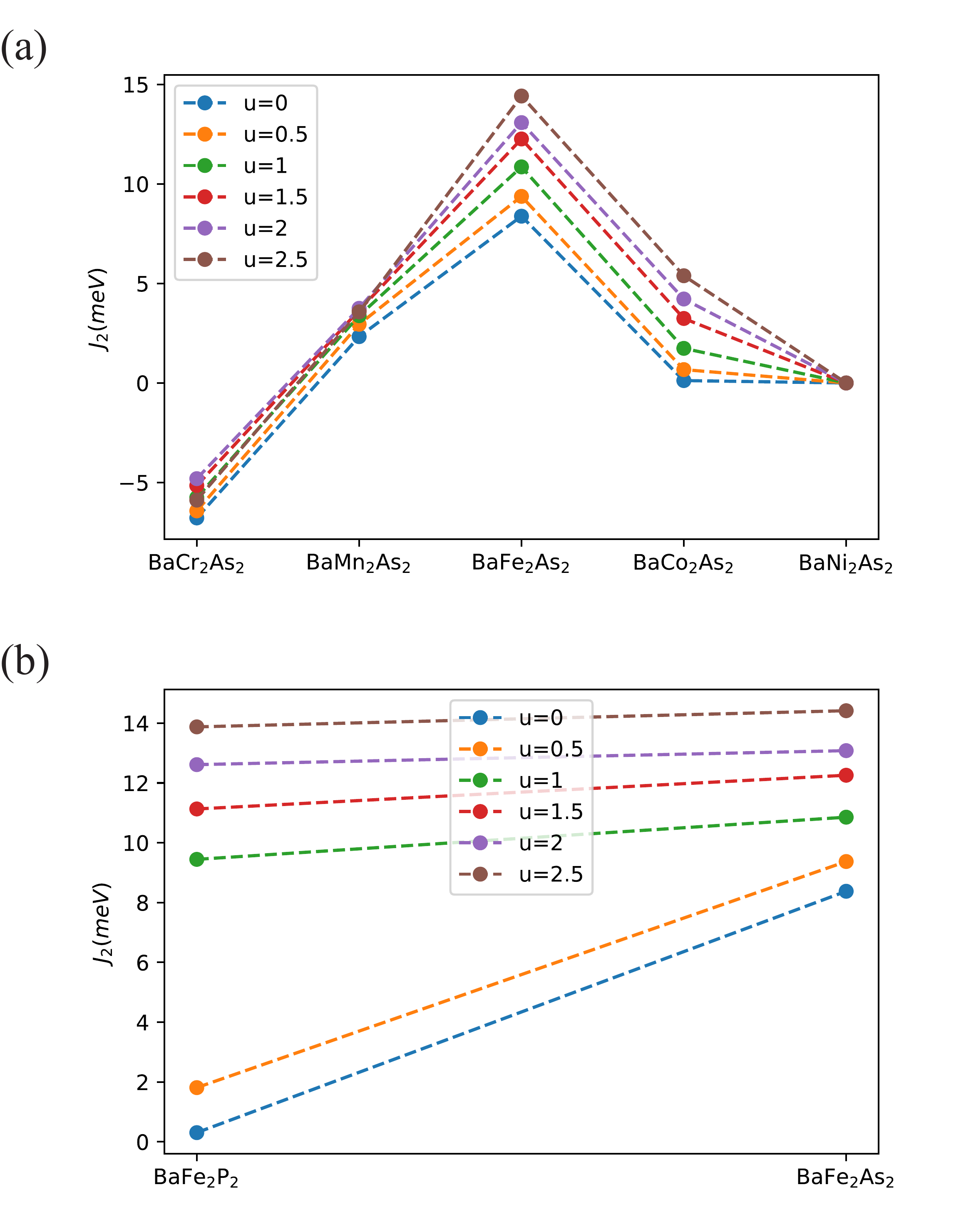}}
\caption{(color online) The $J_2$ magnetic exchange coupling constants of BaM$_2$As$_2$ (M=Cr, Mn, Fe, Co, Ni) and BaFe$_2$B$_2$ (B=P, As).}
\label{tetragonal}
\end{figure}

The existence of large NNN AFM exchange couplings in iron-based materials, namely, $J_2$, is the key difference to separate them from Cr/Mn-based counterparts\cite{jpHusciencebult}.  Differing from the NN exchange couplings, $J_1$, which stems from the direct exchange mechanism,  the $J_2$  exchange couplings are mainly contributed from the superexchange mechanism through the d-p coupling.  In the Fig. \ref{tetragonal}, we plot $J_2$ exchange coupling constants as a function of transition metal elements.  Fig. \ref{tetragonal}(a) is  for  BaM$_2$As$_2$ (M=Cr, Mn, Fe, Co, Ni) and Fig. \ref{tetragonal}(b) is for BaFe$_2$B$_2$(B=P, As). From this result, it is clear that significant AFM $J_2$ only exists in iron-based materials.  This result is a strong support to the high temperature superconductivity in the iron-based superconductors is directly tied with the AFM $J_2$\cite{jpHusciencebult}.

\section{The magnetism in hexagonal 122 transition-metal pnictides}\label{S4}
\begin{figure}[tb]
\centerline{\includegraphics[width=0.5\textwidth]{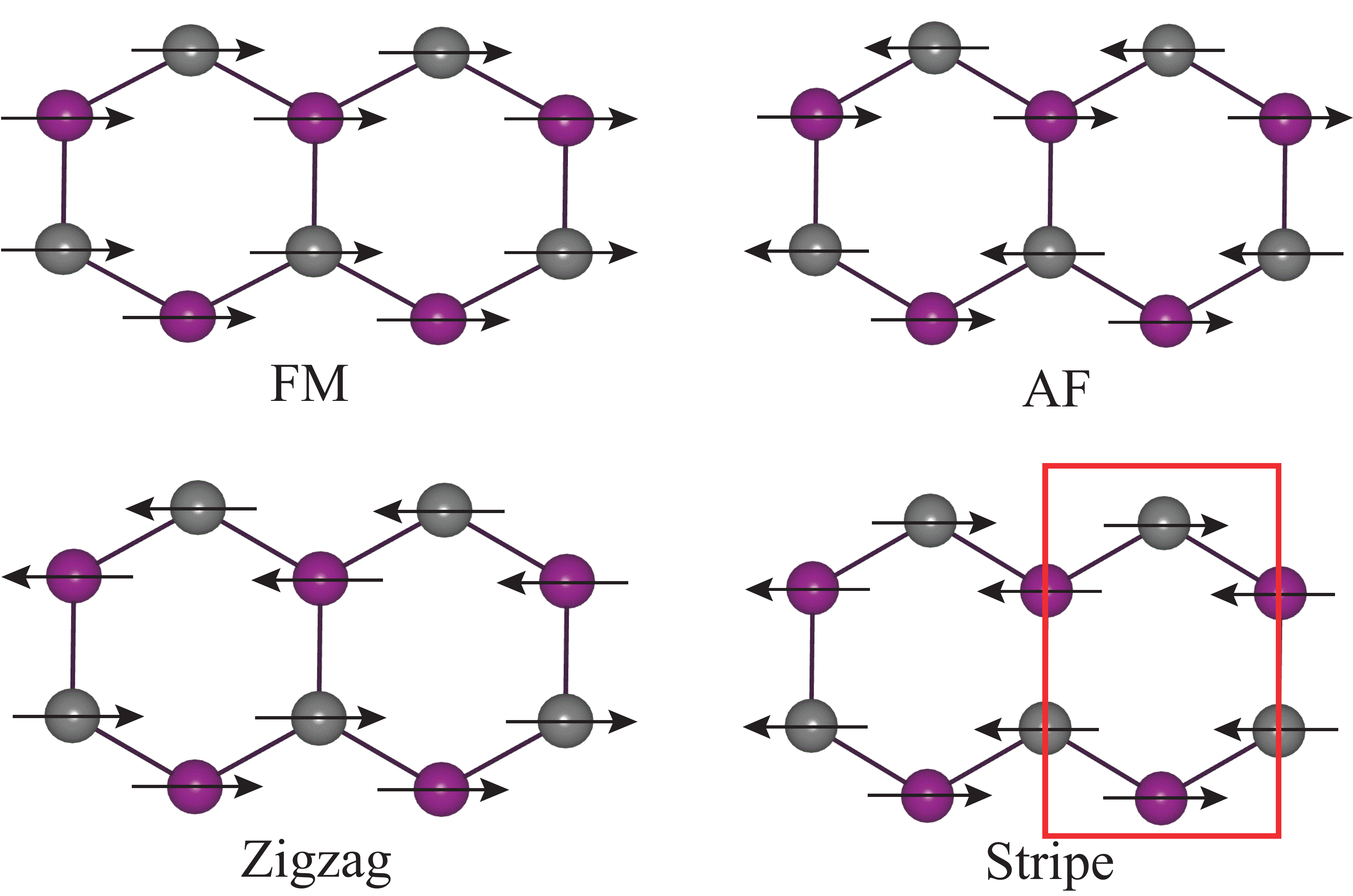}}
\caption{(color online) The sketch of four collinear magnetic states: the FM states, the AFM state, the zigzag state and the stripe state. The red frame indicates the magnetic unit cell in the GGA$+U$ calculations of these magnetic structures.}
\label{magnetic-states}
\end{figure}

 In this section, we use the same method in above section to study the magnetic properties of CaM$_2$B$_2$(M=Cr, Mn, Fe, Co, Ni; B=As, P) with hexagonal structure. We consider four possible collinear magnetic states, the FM state, the AFM state, the zigzag state and the stripe state, which are shown in Fig. \ref{magnetic-states}.  It is also reasonable to assume that the magnetic properties can be approximately described by the $J_1-J_2-J_3$ Heisenberg model\cite{Fang2009, Hu2012u, McNally2015}, which is  well study on the honeycomb lattice\cite{Rastelli1979, Fouet2001, Oitmaa2011}. And the above four collinear magnetic states are contained in the classical phase diagram of the model on the honeycomb lattice. The couplings between the layers, namely $J_z$, is ignored here as it is of a different order of magnitude.
\begin{equation}
\begin{aligned}
\emph{H}=J_{1}\sum_{\langle ij \rangle}S_{i}\cdot S_{j} + J_{2}\sum_{\langle\langle ij \rangle\rangle}S_{i}\cdot S_{j}+J_{3}\sum_{\langle\langle\langle ij\rangle\rangle\rangle}S_{i}\cdot S_{j}
\end{aligned}
\label{hh}
\end{equation}
whereas $\langle ij \rangle$ , $\langle\langle ij \rangle\rangle$ and $\langle\langle\langle ij\rangle\rangle\rangle$ denote the summation over the NN, NNN and third NN, respectively. $S_{i}$ is the spin operator for the $i$th site, $J_{3}$ is the third NN exchange coupling constant.
The classical energies of the above magnetic states are
\begin{eqnarray}
\label{energy-to-magnetic-state-hexagnoal}
E_{FM}   & = & NS^2(6J_{1} + 12J_{2} + 6J_{3})/4 + E_{0} \nonumber\\
E_{AFM}  & = & NS^2(-6J_{1} + 12J_{2} - 6J_{3})/4+ E_{0} \nonumber \\
E_{zigzag} & = & NS^2(2J_{1} - 4J_{2} - 6J_{3})/4+ E_{0}     \nonumber \\
E_{stripe} & = & NS^2(-2J_{1} - 4J_{2} + 6J_{3})/4+ E_{0}
\end{eqnarray}
where $E_0$ is the energy of nonmagnetic state.

We calculated the magnetic moments and the total energies of the above magnetic states with GGA$+U$. If the calculated local magnetic moments on M site are very close to each other, we can extract the magnetic exchange coupling constants by Eqs. (\ref{energy-to-magnetic-state-hexagnoal}). For the S values, we also adopt the high spin values the same in Section \ref{S3}.

\begin{figure}[tb]
\centerline{\includegraphics[width=0.4\textwidth]{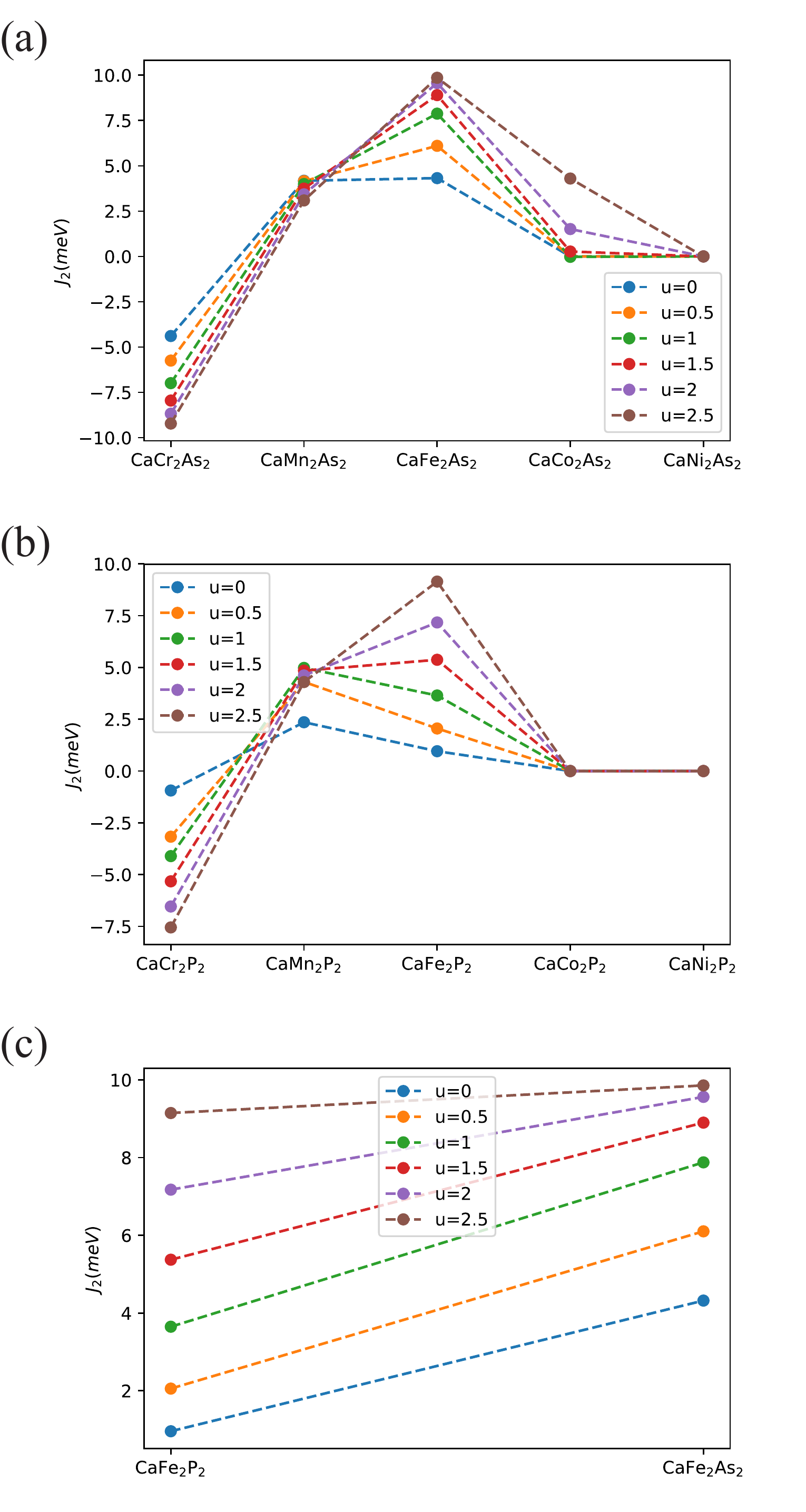}}
\caption{(color online) The J2 exchange coupling parameters in CaM$_2$As$_2$ (M=Cr, Mn, Fe, Co, Ni)(a), CaM$_2$P$_2$ (M=Cr, Mn, Fe, Co, Ni)(b) and CaFe$_2$B$_2$ (B=P, As)(c). }
\label{hexagonal}
\end{figure}

Our calculations suggest that the magnetism in the  hexagonal materials has very similar trend  as those of  the tetragonal counterparts  in the previous section.  In CaCr$_2$As$_2$ the average moments of the four magnetic states range from 3.2 to 3.7 $\mu_{B}$ and the values increases when the $U_{eff}$ increases. The moments vary weakly in the above four magnetic states in each $U_{eff}$. The case in CaCr$_2$P$_2$ is very similar to the case in CaCr$_2$As$_2$, but the moments values are a little smaller than the values in CaCr$_2$As$_2$, in which the average moments of the four magnetic states are in the range $3.0-3.6 \mu_{B}$, except that the average moment is $2.47\mu_{B}$ at $U_{eff}=0$. The average moments at different $U_{eff}$  are in the range $3.7-4.2\mu_{B}$( $3.4-4.1\mu_{B}$) in CaMn$_2$As$_2$ (CaMn$_2$P$_2$) and $2.3-3.1\mu_{B}$($1.6-2.7\mu_{B}$) in CaFe$_2$As$_2$ (CaFe$_2$P$_2$). In CaCo$_2$As$_2$(CaCo$_2$P$_2$) the moments are very small and the moments vary more strongly than the CaCr$_2$As$_2$(CaCr$_2$P$_2$), CaMn$_2$As$_2$(CaMn$_2$P$_2$) and CaFe$_2$As$_2$(CaFe$_2$P$_2$).  Finally, in the CaNi$_2$As$_2$(CaNi$_2$P$_2$) the moments are zero within the range of error $0.005\mu_{B}$ and the energies are nearly degeneracy. Note that, our DFT result show that the AFM state has the lowest energy in CaMn$_2$As$_2$, which is consistent with the experiment result\cite{Sangeetha2016}. Following the same procedure in the previous section,  we can extract the magnetic exchange coupling parameters by Eqs. (\ref{energy-to-magnetic-state-hexagnoal}) quite accurately for  CaCr$_2$As$_2$(CaCr$_2$P$_2$), CaMn$_2$As$_2$(CaCr$_2$P$_2$) and CaFe$_2$As$_2$(CaCr$_2$P$_2$). However,  the calculated exchange parameters are also not accurate in CaCo$_2$As$_2$(CaCo$_2$P$_2$) due to small magnetic moments and the calculated magnetic exchange parameters in CaNi$_2$As$_2$(CaNi$_2$P$_2$) are also zero. The results are summarized in Fig. \ref{hexagonal}.

From these results,  we can also find that  $J_2$  is AFM and reaches the maximum in CaFe$_2$As$_2$ and  it is  small and even ferromagnetic in Cr/Mn-based counterparts.  In CaFe$_2$P$_2$,  $J_2$ exchange is AFM and significant when $U_{eff}\geqslant 1.5$eV. $J_2$  is also larger in CaFe$_2$As$_2$  than  in CaFe$_2$P$_2$ for the same value of $U$ as shown in Fig.\ref{hexagonal}.

In summary, we find that the trend of the magnetism in the hexagonal materials from Cr to Ni-based is very similar to those of the tetragonal counterparts.  The strong AFM exchange coupling between two NNN transition metal atoms only exists in Fe-based materials. As we will show later in this paper, the presence of strong NNN AFM in iron-based materials is consistent with the existence of two near half filling bands that are attributed to the $t_{2g}$ d-orbitals which have large d-p coupling to mediate superexchange AFM couplings.

\section{Pressure Effect on Magnetism}\label{S5}
Both magnetism and superconductivity in the tetragonal iron-based superconductors are known to be sensitive to external pressure\cite{pressure2008, CaFe2As2Pressure, SrBaFe2AS2Pressure}. Here we investigate that the pressure effect on the magnetism of the hexagonal structure CaFe$_2$As$_2$ .

\begin{figure}[tb]
\centerline{\includegraphics[width=0.45\textwidth]{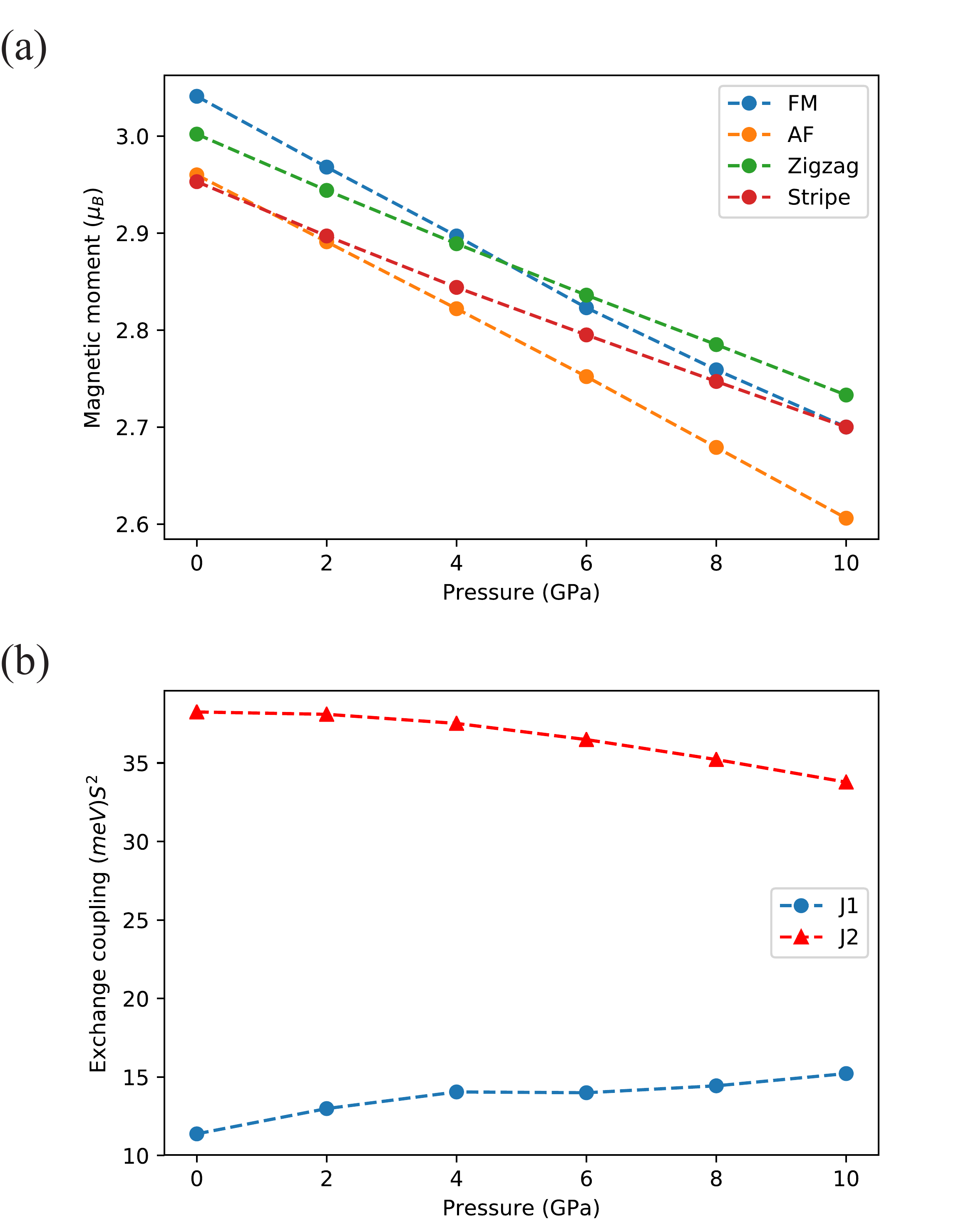}}
\caption{(color online) (a) The magnetic moments as functions of pressure; (b)  the   superexchange antiferromagnetic interaction constants $J_1$ and $J_2$  as a function of the  pressure for CaFe$_2$As$_2$.}
\label{pressure}
\end{figure}

We relax the lattice constants and internal atomic positions with the GGA$+U$ method($U_{eff}=1.5 eV$) in the AFM ordered state under pressure. Then we use the optimized structural parameters to calculate the energies of the above four magnetic states with $U_{eff} = 2$ eV. Using the same method in Section \ref{S4}, we can get the magnetic moments of the above four magnetic states and the exchange coupling parameters. Figure \ref{pressure}(a) shows the pressure dependence of the magnetic moments for FM, AFM, zigzag and stripe states of CaFe$_2$As$_2$.  The magnetic moments in the magnetic states decrease almost linearly with pressure. The magnetic moments are larger than 2$\mu_{B}$ when the pressures are in the range 0-10 Gpa.
Figure \ref{pressure}(b) shows the change of the exchange coupling constants $J_1$ and $J_2$  as a function of the pressure. We can find that both $J_1$ and $J_2$ are very robust against pressure. $J_1$ increases slightly while $J_2$  only slightly decreases under increasing pressure.

These results are similar to to those in the tetragonal  AFe$_2$As$_2$ (A=Ca, Sr, Ba)\cite{Ma2010}. The only difference is that the value of $J_1$ decreases in the latter under increasing pressure.

\section{Electronic Structures on the hexagonal $CaFe_2As_2$}\label{S6}
\begin{figure}[tb]
\centerline{\includegraphics[width=0.45\textwidth]{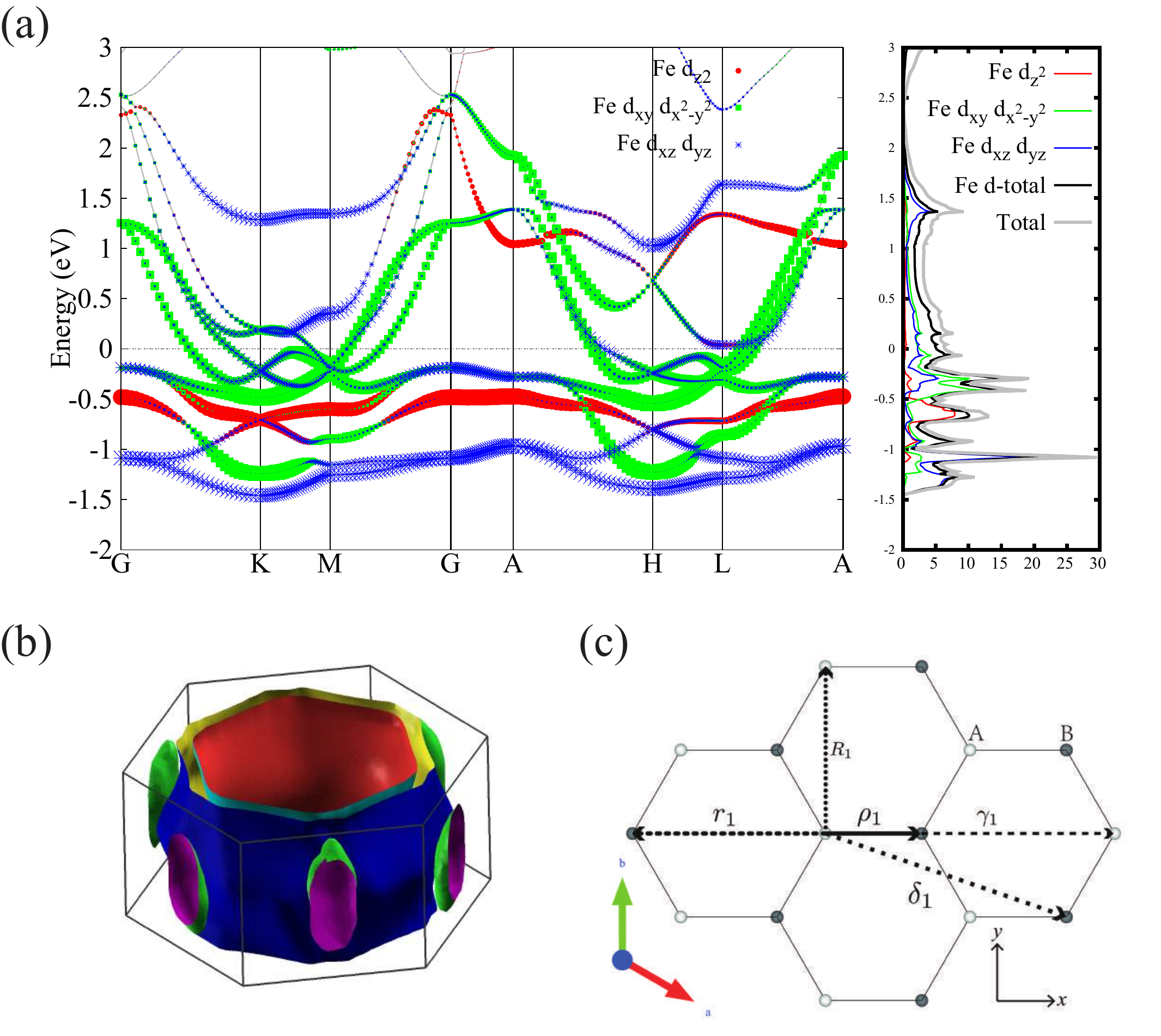}}
\caption{(color online) Band structure, density states(a) and fermi surface(b) of CaFe$_2$As$_2$ with the optimized structural parameters in the paramagnetic state;  (c) the sketch of  the iron lattice:  $\rho_1$ is the NN bond, $R_1$ is the NNN bond, $r_1$ is the third NN bond, $\delta_1$ is the fourth NN bond and $\gamma_1$ is the fifth NN bond. The white dots denote the A sublattice and the gray dots denotes the B sublattice.}
\label{electronic-structure}
\end{figure}

The electronic structures of CaFe$_2$As$_2$ in the paramagnetic state, including  the band structure, density of states (DOS) and fermi surface of CaFe$_2$As$_2$ with the optimized structural parameters, are shown in Fig.\ref{electronic-structure}. As shown in Fig. \ref{electronic-structure}(a, b), there are three Fermi surface sheets, contributed from the three bands crossing the Fermi energy. Among them, the two big quasi-two dimensional cylinder-like Fermi surface sheets centered around $\Gamma$ point are  electron pockets. The remain one centered around $M$ point  forms a small three dimensional Fermi surfaces.  These pockets are attributed to  the $3d$ orbitals of Fe, which are located from -1.5 eV to 2.5 eV as shown in Fig. \ref{electronic-structure}(a).  If we ignore the couplings between FeAs layers along z-direction, the electronic structure simply includes the first two electron pockets.

We can also notice some intriguing features in the distribution of  the five Fe 3d orbitals in the electronic band structure. In the hexagonal lattice structure,  the three $t_{2g}$ d orbitals in the tetrahedron FeAs$_4$, which have higher energy than the e$_g$ orbitals, include $d_{z^2}$ and two other orbitals which are formed by  linear combinations of the other four d-orbitals in which $d_{xy/x^2-y^2}$ carry larger weight than $d_{xz/yz}$.  In Fig. \ref{electronic-structure}(a), we  notice that the DOS of $d_{z^2}$ orbital near the fermi energy is almost zero.  This feature can be understood as follows.  The distance between two NN Fe atoms is very short, which is about 2.902 {\AA}. And the distance between two NNN Fe atoms is about 4.086 {\AA}.  Due to the short NN Fe-Fe distance, the $d_{z^2}$ orbital in the two Fe strongly couples to each other and form two molecular orbitals which can be called as bonding and antibonding orbitals. The bonding orbital is push down below the fermi energy while the antibonding orbital is push up above the fermi energy. This is similar to the $d_{x^2-y^2}$ orbital in tetrahedral iron-based superconductors\cite{jpHuprx, jpHusciencebult}. Near Fermi level, the $d_{xy/x^2-y^2}$ orbitals have the largest weight. This is consistent with the  presence of the large AFM $J_2$  obtained for the iron materials but not others in the previous section because the $d_{xy/x^2-y^2}$ orbitals have larger in-plane coupling to p-orbitals of As than the other orbitals.

We can construct a microscopic electronic model to capture the band structure of  CaFe$_2$As$_2$ by using maximally-localized Wannier orbital calculations\cite{wannier1997, wannier2001}. These maximally localized Wannier functions, centered at the two Fe sites in the unit cell, have five orbital(orbital 1:$d_{3z^2 -r^2}$, 2:$d_{xz}$, 3:$d_{yz}$, 4:$d_{x^2-y^2}$, 5:$d_{xy}$) for each Fe site. Thus,  ten orbitals are needed to describe the tight binding model. As shown in the Fig. \ref{electronic-structure}(a), the band dispersion is very similar between $k_z = 0$ plane and and $k_z = \pi$  plane, except the $d_{z^2}$ orbital above the fermi level. This suggests that the electronic physics is quasi-two dimensional, similar to both cuprates and iron-based superconductors in which the intrinsic interesting physics is known to be two dimensional\cite{jpHuprx, jpHusciencebult, Kurohi2008, rmpPickett1989}. Therefore, for simplicity, we  construct a two dimensional model with the in-plane tight binding couplings,
\begin{eqnarray}
\label{tight-binding-wannier}
H=\sum_{i\vec{\Delta}}\sum_{mn}\sum_{\mu\nu\sigma} t_{i\vec{\Delta},\mu\nu}^{mn} c_{i, m\mu\sigma}^{\dagger} c_{i+\vec{\Delta}, n\nu\sigma} + \sum_{im\mu} (\varepsilon_{\mu}-\mu) n_{im\mu}
\end{eqnarray}

where $t_{i\vec{\Delta},\mu\nu}^{mn}$ are in-plain hoping integrals, $\vec{\Delta}$ is the hopping vector, $m/n = A, B$ labels sublattice, $\mu/\nu = 1,...,5$ labels orbital. $c_{i, m\mu\sigma}^{\dagger}$ creates an electron with spin $\sigma$ on the $\mu$th orbital at site $i$ of $m$th sublattice, and $n_{im\mu} = c_{im\mu}^{\dagger}c_{im\mu}$.  The on-site energies for the five orbitals are $(\varepsilon_1, \varepsilon_2, \varepsilon_3, \varepsilon_4, \varepsilon_5) = (4.103, 3.967, 3.967, 4.077, 4.077)$ eV and the fermi energy
$\mu = 4.116$ eV. Some in-plain hoping integrals $t_{i\vec{\Delta},\mu\nu}^{mn}$ are provided in Table \ref{tab:hoppping-integrals}, other hopping integrals can be obtained by applying symmetry operations according to the point group D$_{3d}$. For the NN bonds and the third NN bonds,  we can apply the symmetries  including  inversion through the bond center, time-reversal as well as C$_2$ rotations. For the fourth NN bonds, there are no C$_2$ rotations. Explicitly,  the  symmetry operations make $t_{\rho_1,\nu\mu}^{AB} = t_{\rho_1,\mu\nu}^{AB}$, $t_{r_1,\nu\mu}^{AB} = t_{r_1,\mu\nu}^{AB}$ and $t_{\delta_1,\nu\mu}^{AB} = t_{\delta_1,\mu\nu}^{AB}$. The hopping integrals of the other bonds can be obtained by applying the C$_3$ rotations along the z direction for  the NN bonds and the third NN bonds. For the fourth NN bonds the addition $\sigma_v$ symmetry operations are needed. The hopping parameters for the other NNN bonds and the firth NN bonds can be also got by applying the  C$_3$ rotations along the z direction and the $\sigma_v$ symmetry operations. We can rotate the whole lattice by a C$_2$ rotation along  z direction, then translate the lattice in the direction with the vector A to B. In this case,  the  sites in the new lattice B locate at the A sites of the original lattice. The the bond direction for the B site is opposite to the bond direction for the A site of original lattice. The operation gives     $t_{R_i,\nu\mu}^{BB} = t_{-R_i,\mu\nu}^{AA}$ and $t_{\gamma_i,\nu\mu}^{BB} = t_{-\gamma_i,\mu\nu}^{AA}$, which the $R_i$ denote all the second NN bonds and $\gamma_i$ denote all the firth NN bonds.

\begin{table}[tb]
\addtolength{\tabcolsep}{4pt}
\caption{ (Unit in eV) A subset of hopping integrals $t_{i\vec{\Delta},\mu\nu}^{mn}$ up to  the fifth NN. $\vec{\Delta}$ is the hopping vector, $m/n = A, B$  are sublattice indices and $\mu/\nu = 1,...,5$ denote orbitals. $\rho_1$ is the NN bond, $R_1$ is the NNN bond, $r_1$ is the third NN bond, $\delta_1$ is the fourth NN bond and $\gamma_1$ is the fifth NN bond. Other hopping integrals can be obtained by applying the symmetry operations as described in the main text.}
\begin{tabular}{ccccccc}
\hline
\hline
%   & Lattice parameters & &  &  Atomic coordinates & \\
%\hline
       & NN                & NNN                 & 3rd NN            & 4th NN           & 5th NN \\
       & $\vec{\Delta} = \rho_1$ & $\vec{\Delta} = R_1$ & $\vec{\Delta} = r_1$ & $\vec{\Delta} = \delta_1$ & $\vec{\Delta} = \gamma_1$ \\
       & A$\rightarrow$ B  & A$\rightarrow$ A      & A $\rightarrow$ B   & A $\rightarrow$ B   & A $\rightarrow$ A  \\
\hline
(1,1)  &-0.33   & 0.10   &-0.02   &  0.00   &-0.00  \\
%\hline
(1,2)  & 0.14   & 0.05   & 0.00   &-0.01    & 0.00       \\
%\hline
(1,3)  & 0       & 0.07   & 0       & 0.01    & 0 \\
%\hline
(1,4)  & 0.15   & 0.03   & 0.01   & 0.02    &-0.01 \\
%\hline
(1,5)  & 0       &-0.07   & 0       & 0.01    & 0 \\
%\hline
(2,2)  &-0.69   &-0.05   & 0.00   &-0.03    &-0.01  \\
%\hline
(2,3)  & 0       &-0.03   & 0       & 0.00    & 0  \\
%\hline
(2,4)  & 0.04   &-0.02   & 0.03   & 0.00    &-0.01  \\
%\hline
(2,5)  & 0       &-0.10   & 0       & 0.03    & 0  \\
%\hline
(3,3)  & 0.16   & 0.05   & 0.07   & 0.00    & 0.00  \\
%\hline
(3,4)  & 0       & 0.10   & 0       &-0.01   & 0  \\
%\hline
(3,5)  &-0.04   & 0.16   &-0.01   & 0.00    & 0.00  \\
%\hline
(4,4)  &-0.05   & 0.14   & 0.01   & 0.01    & 0.02 \\
%\hline
(4,5)  & 0       &-0.11   & 0       & 0.01    & 0 \\
%\hline
(5,5)  & 0.18   & 0.19   & 0.00   & 0.01    & 0.02  \\
\hline
\hline
\end{tabular}
\label{tab:hoppping-integrals}
\end{table}

\section{Possible superconductivity for hexagonal $CaFe_2As_2$}\label{S7}
In this section, we discuss possible superconducting states  in the hexagonal iron-based materials under the assumption that it is the superexchange couplings to cause superconductivity.

Since the crystal structure of CaFe$_2$As$_2$ belongs to the point group of D$_{3d}$, the pairing symmetry of CaFe$_2$As$_2$ can be classified according to the irreducible representations of the D$_{3d}$ point group.  Moreover, only even parity spin-singlet pairing is allowed if the superconductivity is driven by the AFM exchange couplings. In this case, there  are two possible superconducting states with  $A_{1g}$ (extended s-wave) and $E_g$ (d-wave).  For the $E_g$  d-wave states, there are two degenerate states. The superconducting condensation energy can be further lowered by forming the  time-reversal symmetry breaking d$\pm$id-wave states. Thus we just need to focus on comparing the energies between the extended s-wave and the d$\pm$id-wave states.

\begin{figure}[t]
\centerline{\includegraphics[width=0.5\textwidth]{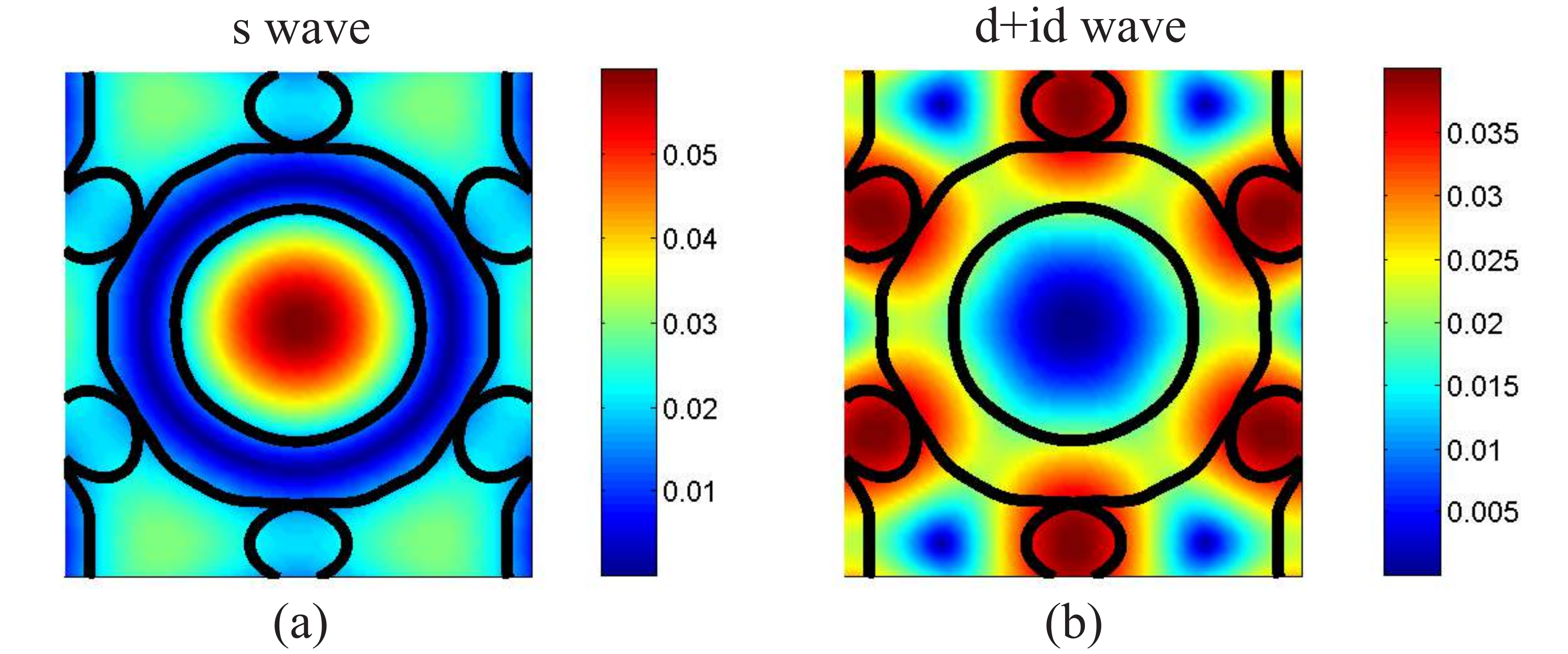}}
\caption{(color online) The overlap between the Fermi Surfaces (black lines) and the  superconducting gap distribution for CaFe$_2$As$_2$ in the first Brillouin zone (BZ) in the $k_z=0$ plane for the extended s-wave (a) and d+id-wave (b) cases.   The s-wave order parameter in $k$-space has a momentum form factor  $\Delta(\bm{k})=\Delta(2\cos\frac{\sqrt{3}}{2}k_x\cos\frac{1}{2}k_y+\cos k_y)$, and the d+id-wave order parameter  has $\Delta(\bm{k})=\Delta(\cos k_y-\cos\frac{\sqrt{3}}{2}k_x\cos\frac{1}{2}k_y+i\sqrt{3}\sin\frac{\sqrt{3}}{2}k_x\sin\frac{1}{2}k_y)$. The color bar labels the amplitude of the SC gap when $\Delta=0.02$.  }
\label{pair}
\end{figure}

A selection rule to determine the superconducting state, which we refer as Hu-Ding principle,  has been proposed\cite{Hu_Ding, Davis29102013} to unify  the d-wave pairing in cuprates and s-wave pairing in iron-based superconductors. The principle states that in order to generating high T$_c$ superconductivity, the momentum space form factor of the superconducting pairing gap function which is determined by the AFM superexchange couplings must have large overlap with Fermi surfaces. The most favored pairing symmetry is the one which has the largest overlap strength\cite{Hu_Ding}. The overlap strength can be defined as

\begin{equation}
\begin{aligned}
W=\int\int dk_xdk_y|\Delta_{\bm{k}}|^2\delta(\epsilon_{\bm{k}}-\mu),
\end{aligned}
\label{overlap}
\end{equation}

where $\Delta_{\bm{k}}$ is the  momentum space SC gap function. In our case,  the gap function stems from the NNN AFM superexchange couplings. For the extended s-wave,  $\Delta(\bm{k})=\Delta(2\cos\frac{\sqrt{3}}{2}k_x\cos\frac{1}{2}k_y+\cos k_y)$, and the d+id-wave order parameter $\Delta(\bm{k})=\Delta(\cos k_y-\cos\frac{\sqrt{3}}{2}k_x\cos\frac{1}{2}k_y+i\sqrt{3}\sin\frac{\sqrt{3}}{2}k_x\sin\frac{1}{2}k_y)$, where    $\Delta$ is a constant. In Fig. \ref{pair},  we plot the overlap strength   for both  the s-wave Fig. \ref{pair}(a) and d+id-wave Fig. \ref{pair}(b) cases for 0.2 electron doping.  The overlap strength for the d$\pm$id-wave is $2.85$ times larger than the s-wave case (with an energy cutoff $0.05 eV$ from the Fermi energy), so the most favored pairing symmetry for the hexagonal CaFe$_2$As$_2$ is the d+id-wave.  Therefore, the d$\pm$id superconducting state is favored.

\section{ Discussion and Conclusion}\label{S8}
In summary, we have shown that the hexagonal transition metal pnictides have very similar trend in magnetic exchange interactions  as those tetragonal counterparts. In both cases, the iron-based materials maximize the NNN antiferromagnetic interactions and those d-orbitals which are responsible for the largest superexchange  interactions dominate near Fermi surfaces.  The superexchange interactions make the hexagonal iron materials as extremely magnetic frustrated systems and can also  lead to  d+id superconducting ground states upon doping. As the energy scales of the NNN AFM superexchange couplings in both hexagonal and tetragonal iron materials are close to each other, we expect that the hexagonal materials can host high $T_c$ superconductivity, just like the tetragonal counterparts.

Although the hexagonal Mn-based pnictides have been successfully synthesized, the  iron-based counterparts have not been obtained. However, it is worth to mention that the iron-based materials are  stable in our theoretical investigation. Their phonon spectra do not show any imaginary modes.   Our study can be extended to include transition metal chalcogenides.   The similar results  can be expected for those chalcogenides with the similar  hexagonal structures.

\section{Acknowledgements}
 We thank X. X. Wu for helpful discussion. The work is supported by the Ministry of Science and Technology of China 973 program (No. 2015CB921300, No.~2017YFA0303100), National Science Foundation of China (Grant No. NSFC-1190020, 11534014, 11334012), and the Strategic Priority Research Program of CAS (Grant No.XDB07000000). QHW also acknowledges the supports by the NSFC funding No.11574134.

%\bibliography{hexagonalSC-new}

\begin{thebibliography}{68}
\expandafter\ifx\csname natexlab\endcsname\relax\def\natexlab#1{#1}\fi
\expandafter\ifx\csname bibnamefont\endcsname\relax
  \def\bibnamefont#1{#1}\fi
\expandafter\ifx\csname bibfnamefont\endcsname\relax
  \def\bibfnamefont#1{#1}\fi
\expandafter\ifx\csname citenamefont\endcsname\relax
  \def\citenamefont#1{#1}\fi
\expandafter\ifx\csname url\endcsname\relax
  \def\url#1{\texttt{#1}}\fi
\expandafter\ifx\csname urlprefix\endcsname\relax\def\urlprefix{URL }\fi
\providecommand{\bibinfo}[2]{#2}
\providecommand{\eprint}[2][]{\url{#2}}

\bibitem[{\citenamefont{Kamihara et~al.}(2008)\citenamefont{Kamihara, Watanabe,
  Hirano, and Hosono}}]{Fe}
\bibinfo{author}{\bibfnamefont{Y.}~\bibnamefont{Kamihara}},
  \bibinfo{author}{\bibfnamefont{T.}~\bibnamefont{Watanabe}},
  \bibinfo{author}{\bibfnamefont{M.}~\bibnamefont{Hirano}}, \bibnamefont{and}
  \bibinfo{author}{\bibfnamefont{H.}~\bibnamefont{Hosono}},
  \bibinfo{journal}{JACS} \textbf{\bibinfo{volume}{130}}, \bibinfo{pages}{3296}
  (\bibinfo{year}{2008}).

\bibitem[{\citenamefont{Hirschfeld et~al.}(2011)\citenamefont{Hirschfeld,
  Korshunov, and Mazin}}]{Hirschfeld2011}
\bibinfo{author}{\bibfnamefont{P.~J.} \bibnamefont{Hirschfeld}},
  \bibinfo{author}{\bibfnamefont{M.~M.} \bibnamefont{Korshunov}},
  \bibnamefont{and} \bibinfo{author}{\bibfnamefont{I.~I.} \bibnamefont{Mazin}},
  \bibinfo{journal}{Rep. Prog. Phys.} \textbf{\bibinfo{volume}{74}},
  \bibinfo{pages}{4508} (\bibinfo{year}{2011}).

\bibitem[{\citenamefont{Bednorz and Muller}(1986)}]{Cu}
\bibinfo{author}{\bibfnamefont{J.~G.} \bibnamefont{Bednorz}} \bibnamefont{and}
  \bibinfo{author}{\bibfnamefont{K.~A.} \bibnamefont{Muller}},
  \bibinfo{journal}{Z. Phys. B} \textbf{\bibinfo{volume}{64}},
  \bibinfo{pages}{189} (\bibinfo{year}{1986}).

\bibitem[{\citenamefont{Sefat et~al.}(2009{\natexlab{a}})\citenamefont{Sefat,
  Singh, VanBebber, Mozharivskyj, McGuire, Jin, Sales, Keppens, and
  Mandrus}}]{Sefat2009-Cr}
\bibinfo{author}{\bibfnamefont{A.~S.} \bibnamefont{Sefat}},
  \bibinfo{author}{\bibfnamefont{D.~J.} \bibnamefont{Singh}},
  \bibinfo{author}{\bibfnamefont{L.~H.} \bibnamefont{VanBebber}},
  \bibinfo{author}{\bibfnamefont{Y.}~\bibnamefont{Mozharivskyj}},
  \bibinfo{author}{\bibfnamefont{M.~A.} \bibnamefont{McGuire}},
  \bibinfo{author}{\bibfnamefont{R.}~\bibnamefont{Jin}},
  \bibinfo{author}{\bibfnamefont{B.~C.} \bibnamefont{Sales}},
  \bibinfo{author}{\bibfnamefont{V.}~\bibnamefont{Keppens}}, \bibnamefont{and}
  \bibinfo{author}{\bibfnamefont{D.}~\bibnamefont{Mandrus}},
  \bibinfo{journal}{Phys. Rev. B} \textbf{\bibinfo{volume}{79}},
  \bibinfo{pages}{224524} (\bibinfo{year}{2009}{\natexlab{a}}).

\bibitem[{\citenamefont{Kasinathan et~al.}(2009)\citenamefont{Kasinathan,
  Ormeci, Koch, Burkhardt, Schnelle, Leithe-Jasper, and
  Rosner}}]{Kasinathan2009-Mn}
\bibinfo{author}{\bibfnamefont{D.}~\bibnamefont{Kasinathan}},
  \bibinfo{author}{\bibfnamefont{A.}~\bibnamefont{Ormeci}},
  \bibinfo{author}{\bibfnamefont{K.}~\bibnamefont{Koch}},
  \bibinfo{author}{\bibfnamefont{U.}~\bibnamefont{Burkhardt}},
  \bibinfo{author}{\bibfnamefont{W.}~\bibnamefont{Schnelle}},
  \bibinfo{author}{\bibfnamefont{A.}~\bibnamefont{Leithe-Jasper}},
  \bibnamefont{and} \bibinfo{author}{\bibfnamefont{H.}~\bibnamefont{Rosner}},
  \bibinfo{journal}{New Journal of Physics} \textbf{\bibinfo{volume}{11}},
  \bibinfo{pages}{025023} (\bibinfo{year}{2009}).

\bibitem[{\citenamefont{Pandey et~al.}(2011)\citenamefont{Pandey, Anand, and
  Johnston}}]{Pandey2011-Mn}
\bibinfo{author}{\bibfnamefont{A.}~\bibnamefont{Pandey}},
  \bibinfo{author}{\bibfnamefont{V.~K.} \bibnamefont{Anand}}, \bibnamefont{and}
  \bibinfo{author}{\bibfnamefont{D.~C.} \bibnamefont{Johnston}},
  \bibinfo{journal}{Phys. Rev. B} \textbf{\bibinfo{volume}{84}},
  \bibinfo{pages}{014405} (\bibinfo{year}{2011}).

\bibitem[{\citenamefont{Pandey et~al.}(2012)\citenamefont{Pandey, Dhaka,
  Lamsal, Lee, Anand, Kreyssig, Heitmann, McQueeney, Goldman, Harmon
  et~al.}}]{Pandey2012prl-Mn}
\bibinfo{author}{\bibfnamefont{A.}~\bibnamefont{Pandey}},
  \bibinfo{author}{\bibfnamefont{R.~S.} \bibnamefont{Dhaka}},
  \bibinfo{author}{\bibfnamefont{J.}~\bibnamefont{Lamsal}},
  \bibinfo{author}{\bibfnamefont{Y.}~\bibnamefont{Lee}},
  \bibinfo{author}{\bibfnamefont{V.~K.} \bibnamefont{Anand}},
  \bibinfo{author}{\bibfnamefont{A.}~\bibnamefont{Kreyssig}},
  \bibinfo{author}{\bibfnamefont{T.~W.} \bibnamefont{Heitmann}},
  \bibinfo{author}{\bibfnamefont{R.~J.} \bibnamefont{McQueeney}},
  \bibinfo{author}{\bibfnamefont{A.~I.} \bibnamefont{Goldman}},
  \bibinfo{author}{\bibfnamefont{B.~N.} \bibnamefont{Harmon}},
  \bibnamefont{et~al.}, \bibinfo{journal}{Phys. Rev. Lett.}
  \textbf{\bibinfo{volume}{108}}, \bibinfo{pages}{087005}
  (\bibinfo{year}{2012}).

\bibitem[{\citenamefont{Ahilan et~al.}(2014)\citenamefont{Ahilan, Imai, Sefat,
  and Ning}}]{Ahilan2014-Co}
\bibinfo{author}{\bibfnamefont{K.}~\bibnamefont{Ahilan}},
  \bibinfo{author}{\bibfnamefont{T.}~\bibnamefont{Imai}},
  \bibinfo{author}{\bibfnamefont{A.~S.} \bibnamefont{Sefat}}, \bibnamefont{and}
  \bibinfo{author}{\bibfnamefont{F.~L.} \bibnamefont{Ning}},
  \bibinfo{journal}{Phys. Rev. B} \textbf{\bibinfo{volume}{90}},
  \bibinfo{pages}{014520} (\bibinfo{year}{2014}).

\bibitem[{\citenamefont{Anand et~al.}(2014)\citenamefont{Anand, Quirinale, Lee,
  Harmon, Furukawa, Ogloblichev, Huq, Abernathy, Stephens, McQueeney
  et~al.}}]{Anand2014-Co}
\bibinfo{author}{\bibfnamefont{V.~K.} \bibnamefont{Anand}},
  \bibinfo{author}{\bibfnamefont{D.~G.} \bibnamefont{Quirinale}},
  \bibinfo{author}{\bibfnamefont{Y.}~\bibnamefont{Lee}},
  \bibinfo{author}{\bibfnamefont{B.~N.} \bibnamefont{Harmon}},
  \bibinfo{author}{\bibfnamefont{Y.}~\bibnamefont{Furukawa}},
  \bibinfo{author}{\bibfnamefont{V.~V.} \bibnamefont{Ogloblichev}},
  \bibinfo{author}{\bibfnamefont{A.}~\bibnamefont{Huq}},
  \bibinfo{author}{\bibfnamefont{D.~L.} \bibnamefont{Abernathy}},
  \bibinfo{author}{\bibfnamefont{P.~W.} \bibnamefont{Stephens}},
  \bibinfo{author}{\bibfnamefont{R.~J.} \bibnamefont{McQueeney}},
  \bibnamefont{et~al.}, \bibinfo{journal}{Phys. Rev. B}
  \textbf{\bibinfo{volume}{90}}, \bibinfo{pages}{064517}
  (\bibinfo{year}{2014}).

\bibitem[{\citenamefont{Ronning
  et~al.}(2008{\natexlab{a}})\citenamefont{Ronning, Kurita, Bauer, Scott, Park,
  Klimczuk, Movshovich, and Thompson}}]{Ronning2008-Ni}
\bibinfo{author}{\bibfnamefont{F.}~\bibnamefont{Ronning}},
  \bibinfo{author}{\bibfnamefont{N.}~\bibnamefont{Kurita}},
  \bibinfo{author}{\bibfnamefont{E.~D.} \bibnamefont{Bauer}},
  \bibinfo{author}{\bibfnamefont{B.~L.} \bibnamefont{Scott}},
  \bibinfo{author}{\bibfnamefont{T.}~\bibnamefont{Park}},
  \bibinfo{author}{\bibfnamefont{T.}~\bibnamefont{Klimczuk}},
  \bibinfo{author}{\bibfnamefont{R.}~\bibnamefont{Movshovich}},
  \bibnamefont{and} \bibinfo{author}{\bibfnamefont{J.~D.}
  \bibnamefont{Thompson}}, \bibinfo{journal}{Journal of Physics: Condensed
  Matter} \textbf{\bibinfo{volume}{20}}, \bibinfo{pages}{342203}
  (\bibinfo{year}{2008}{\natexlab{a}}).

\bibitem[{\citenamefont{Zhang and Zhai}(2017)}]{ZhangPan2017-Ni}
\bibinfo{author}{\bibfnamefont{P.}~\bibnamefont{Zhang}} \bibnamefont{and}
  \bibinfo{author}{\bibfnamefont{H.-f.} \bibnamefont{Zhai}},
  \bibinfo{journal}{Condensed Matter} \textbf{\bibinfo{volume}{2}},
  \bibinfo{pages}{28} (\bibinfo{year}{2017}).

\bibitem[{\citenamefont{Saparov and Sefat}(2012)}]{Saparov2012-Cu}
\bibinfo{author}{\bibfnamefont{B.}~\bibnamefont{Saparov}} \bibnamefont{and}
  \bibinfo{author}{\bibfnamefont{A.~S.} \bibnamefont{Sefat}},
  \bibinfo{journal}{Journal of Solid State Chemistry}
  \textbf{\bibinfo{volume}{191}}, \bibinfo{pages}{213 } (\bibinfo{year}{2012}).

\bibitem[{\citenamefont{Anand et~al.}(2012)\citenamefont{Anand, Perera, Pandey,
  Goetsch, Kreyssig, and Johnston}}]{Anand2012-Cu}
\bibinfo{author}{\bibfnamefont{V.~K.} \bibnamefont{Anand}},
  \bibinfo{author}{\bibfnamefont{P.~K.} \bibnamefont{Perera}},
  \bibinfo{author}{\bibfnamefont{A.}~\bibnamefont{Pandey}},
  \bibinfo{author}{\bibfnamefont{R.~J.} \bibnamefont{Goetsch}},
  \bibinfo{author}{\bibfnamefont{A.}~\bibnamefont{Kreyssig}}, \bibnamefont{and}
  \bibinfo{author}{\bibfnamefont{D.~C.} \bibnamefont{Johnston}},
  \bibinfo{journal}{Phys. Rev. B} \textbf{\bibinfo{volume}{85}},
  \bibinfo{pages}{214523} (\bibinfo{year}{2012}).

\bibitem[{\citenamefont{Dai}(2015)}]{DaiPengchengRMP2015}
\bibinfo{author}{\bibfnamefont{P.}~\bibnamefont{Dai}}, \bibinfo{journal}{Rev.
  Mod. Phys.} \textbf{\bibinfo{volume}{87}}, \bibinfo{pages}{855}
  (\bibinfo{year}{2015}).

\bibitem[{\citenamefont{Seo et~al.}(2008)\citenamefont{Seo, Bernevig, and
  Hu}}]{Seo2008}
\bibinfo{author}{\bibfnamefont{K.~J.} \bibnamefont{Seo}},
  \bibinfo{author}{\bibfnamefont{B.~A.} \bibnamefont{Bernevig}},
  \bibnamefont{and} \bibinfo{author}{\bibfnamefont{J.~P.} \bibnamefont{Hu}},
  \bibinfo{journal}{Phys. Rev. Lett.} \textbf{\bibinfo{volume}{101}},
  \bibinfo{pages}{206404} (\bibinfo{year}{2008}).

\bibitem[{\citenamefont{Fang et~al.}(2011)\citenamefont{Fang, Wu, Thomale,
  Bernevig, and Hu}}]{Fang2011}
\bibinfo{author}{\bibfnamefont{C.}~\bibnamefont{Fang}},
  \bibinfo{author}{\bibfnamefont{Y.~L.} \bibnamefont{Wu}},
  \bibinfo{author}{\bibfnamefont{R.}~\bibnamefont{Thomale}},
  \bibinfo{author}{\bibfnamefont{B.~A.} \bibnamefont{Bernevig}},
  \bibnamefont{and} \bibinfo{author}{\bibfnamefont{J.~P.} \bibnamefont{Hu}},
  \bibinfo{journal}{Phys. Rev. X} \textbf{\bibinfo{volume}{1}}
  (\bibinfo{year}{2011}).

\bibitem[{\citenamefont{Hu and Yuan}(2016)}]{Hu-Yuan}
\bibinfo{author}{\bibfnamefont{J.}~\bibnamefont{Hu}} \bibnamefont{and}
  \bibinfo{author}{\bibfnamefont{J.}~\bibnamefont{Yuan}},
  \bibinfo{journal}{Frontiers of Physics} \textbf{\bibinfo{volume}{11}},
  \bibinfo{pages}{117404} (\bibinfo{year}{2016}).

\bibitem[{\citenamefont{Hu et~al.}(2015)\citenamefont{Hu, Le, and
  Wu}}]{jpHuprx}
\bibinfo{author}{\bibfnamefont{J.}~\bibnamefont{Hu}},
  \bibinfo{author}{\bibfnamefont{C.}~\bibnamefont{Le}}, \bibnamefont{and}
  \bibinfo{author}{\bibfnamefont{X.}~\bibnamefont{Wu}}, \bibinfo{journal}{Phys.
  Rev. X} \textbf{\bibinfo{volume}{5}}, \bibinfo{pages}{041012}
  (\bibinfo{year}{2015}).

\bibitem[{\citenamefont{Hu}(2016)}]{jpHusciencebult}
\bibinfo{author}{\bibfnamefont{J.}~\bibnamefont{Hu}}, \bibinfo{journal}{Science
  Bulletin} \textbf{\bibinfo{volume}{61}}, \bibinfo{pages}{561}
  (\bibinfo{year}{2016}).

\bibitem[{\citenamefont{Kresse and Hafner}(1993)}]{Kresse1993}
\bibinfo{author}{\bibfnamefont{G.}~\bibnamefont{Kresse}} \bibnamefont{and}
  \bibinfo{author}{\bibfnamefont{J.}~\bibnamefont{Hafner}},
  \bibinfo{journal}{Phys. Rev. B} \textbf{\bibinfo{volume}{47}},
  \bibinfo{pages}{558} (\bibinfo{year}{1993}).

\bibitem[{\citenamefont{Kresse and Furthm\"uller}(1996)}]{Kresse1996}
\bibinfo{author}{\bibfnamefont{G.}~\bibnamefont{Kresse}} \bibnamefont{and}
  \bibinfo{author}{\bibfnamefont{J.}~\bibnamefont{Furthm\"uller}},
  \bibinfo{journal}{Computational Materials Science}
  \textbf{\bibinfo{volume}{6}}, \bibinfo{pages}{15 } (\bibinfo{year}{1996}).

\bibitem[{\citenamefont{Kresse and Furthm\"uller}(1996)}]{Kresse1996B}
\bibinfo{author}{\bibfnamefont{G.}~\bibnamefont{Kresse}} \bibnamefont{and}
  \bibinfo{author}{\bibfnamefont{J.}~\bibnamefont{Furthm\"uller}},
  \bibinfo{journal}{Phys. Rev. B} \textbf{\bibinfo{volume}{54}},
  \bibinfo{pages}{11169} (\bibinfo{year}{1996}).

\bibitem[{\citenamefont{Perdew et~al.}(1997)\citenamefont{Perdew, Burke, and
  Ernzerhof}}]{Perdew1996}
\bibinfo{author}{\bibfnamefont{J.~P.} \bibnamefont{Perdew}},
  \bibinfo{author}{\bibfnamefont{K.}~\bibnamefont{Burke}}, \bibnamefont{and}
  \bibinfo{author}{\bibfnamefont{M.}~\bibnamefont{Ernzerhof}},
  \bibinfo{journal}{Physical Review Letters} \textbf{\bibinfo{volume}{78}},
  \bibinfo{pages}{1396} (\bibinfo{year}{1997}).

\bibitem[{\citenamefont{Dudarev et~al.}(1998)\citenamefont{Dudarev, Botton,
  Savrasov, Humphreys, and Sutton}}]{Dudarev1998}
\bibinfo{author}{\bibfnamefont{S.~L.} \bibnamefont{Dudarev}},
  \bibinfo{author}{\bibfnamefont{G.~A.} \bibnamefont{Botton}},
  \bibinfo{author}{\bibfnamefont{S.~Y.} \bibnamefont{Savrasov}},
  \bibinfo{author}{\bibfnamefont{C.~J.} \bibnamefont{Humphreys}},
  \bibnamefont{and} \bibinfo{author}{\bibfnamefont{A.~P.}
  \bibnamefont{Sutton}}, \bibinfo{journal}{Physical Review B}
  \textbf{\bibinfo{volume}{57}}, \bibinfo{pages}{1505} (\bibinfo{year}{1998}).

\bibitem[{\citenamefont{Pfisterer and Nagorsen}(1980)}]{BaCr2As2andBaCo2As2}
\bibinfo{author}{\bibfnamefont{M.}~\bibnamefont{Pfisterer}} \bibnamefont{and}
  \bibinfo{author}{\bibfnamefont{G.}~\bibnamefont{Nagorsen}},
  \bibinfo{journal}{Zeitschrift fuer Naturforschung}
  \textbf{\bibinfo{volume}{35b}}, \bibinfo{pages}{703} (\bibinfo{year}{1980}).

\bibitem[{\citenamefont{Singh et~al.}(2009{\natexlab{a}})\citenamefont{Singh,
  Green, Huang, Kreyssig, McQueeney, Johnston, and Goldman}}]{BaMn2As2}
\bibinfo{author}{\bibfnamefont{Y.}~\bibnamefont{Singh}},
  \bibinfo{author}{\bibfnamefont{M.~A.} \bibnamefont{Green}},
  \bibinfo{author}{\bibfnamefont{Q.}~\bibnamefont{Huang}},
  \bibinfo{author}{\bibfnamefont{A.}~\bibnamefont{Kreyssig}},
  \bibinfo{author}{\bibfnamefont{R.~J.} \bibnamefont{McQueeney}},
  \bibinfo{author}{\bibfnamefont{D.~C.} \bibnamefont{Johnston}},
  \bibnamefont{and} \bibinfo{author}{\bibfnamefont{A.~I.}
  \bibnamefont{Goldman}}, \bibinfo{journal}{Phys. Rev. B}
  \textbf{\bibinfo{volume}{80}}, \bibinfo{pages}{100403}
  (\bibinfo{year}{2009}{\natexlab{a}}).

\bibitem[{\citenamefont{Rotter et~al.}(2008)\citenamefont{Rotter, Tegel,
  Johrendt, Schellenberg, Hermes, and P\"ottgen}}]{BaFe2As2}
\bibinfo{author}{\bibfnamefont{M.}~\bibnamefont{Rotter}},
  \bibinfo{author}{\bibfnamefont{M.}~\bibnamefont{Tegel}},
  \bibinfo{author}{\bibfnamefont{D.}~\bibnamefont{Johrendt}},
  \bibinfo{author}{\bibfnamefont{I.}~\bibnamefont{Schellenberg}},
  \bibinfo{author}{\bibfnamefont{W.}~\bibnamefont{Hermes}}, \bibnamefont{and}
  \bibinfo{author}{\bibfnamefont{R.}~\bibnamefont{P\"ottgen}},
  \bibinfo{journal}{Phys. Rev. B} \textbf{\bibinfo{volume}{78}},
  \bibinfo{pages}{020503} (\bibinfo{year}{2008}).

\bibitem[{\citenamefont{Ronning
  et~al.}(2008{\natexlab{b}})\citenamefont{Ronning, Kurita, Bauer, Scott, Park,
  Klimczuk, Movshovich, and Thompson}}]{BaNi2As2}
\bibinfo{author}{\bibfnamefont{F.}~\bibnamefont{Ronning}},
  \bibinfo{author}{\bibfnamefont{N.}~\bibnamefont{Kurita}},
  \bibinfo{author}{\bibfnamefont{E.~D.} \bibnamefont{Bauer}},
  \bibinfo{author}{\bibfnamefont{B.~L.} \bibnamefont{Scott}},
  \bibinfo{author}{\bibfnamefont{T.}~\bibnamefont{Park}},
  \bibinfo{author}{\bibfnamefont{T.}~\bibnamefont{Klimczuk}},
  \bibinfo{author}{\bibfnamefont{R.}~\bibnamefont{Movshovich}},
  \bibnamefont{and} \bibinfo{author}{\bibfnamefont{J.~D.}
  \bibnamefont{Thompson}}, \bibinfo{journal}{Journal of Physics: Condensed
  Matter} \textbf{\bibinfo{volume}{20}}, \bibinfo{pages}{342203}
  (\bibinfo{year}{2008}{\natexlab{b}}).

\bibitem[{\citenamefont{Arnold et~al.}(2011)\citenamefont{Arnold, Kasahara,
  Coldea, Terashima, Matsuda, Shibauchi, and Carrington}}]{BaFe2P2}
\bibinfo{author}{\bibfnamefont{B.~J.} \bibnamefont{Arnold}},
  \bibinfo{author}{\bibfnamefont{S.}~\bibnamefont{Kasahara}},
  \bibinfo{author}{\bibfnamefont{A.~I.} \bibnamefont{Coldea}},
  \bibinfo{author}{\bibfnamefont{T.}~\bibnamefont{Terashima}},
  \bibinfo{author}{\bibfnamefont{Y.}~\bibnamefont{Matsuda}},
  \bibinfo{author}{\bibfnamefont{T.}~\bibnamefont{Shibauchi}},
  \bibnamefont{and}
  \bibinfo{author}{\bibfnamefont{A.}~\bibnamefont{Carrington}},
  \bibinfo{journal}{Phys. Rev. B} \textbf{\bibinfo{volume}{83}},
  \bibinfo{pages}{220504} (\bibinfo{year}{2011}).

\bibitem[{\citenamefont{Brechtel et~al.}(1978)\citenamefont{Brechtel, Cordier,
  and Schaefer}}]{CaMn2As2}
\bibinfo{author}{\bibfnamefont{E.}~\bibnamefont{Brechtel}},
  \bibinfo{author}{\bibfnamefont{G.}~\bibnamefont{Cordier}}, \bibnamefont{and}
  \bibinfo{author}{\bibfnamefont{H.}~\bibnamefont{Schaefer}},
  \bibinfo{journal}{Zeitschrift fuer Naturforschung}
  \textbf{\bibinfo{volume}{33b}}, \bibinfo{pages}{820} (\bibinfo{year}{1978}).

\bibitem[{\citenamefont{Mewis}(1978)}]{CaMn2P2}
\bibinfo{author}{\bibfnamefont{A.}~\bibnamefont{Mewis}},
  \bibinfo{journal}{Zeitschrift fuer Naturforschung}
  \textbf{\bibinfo{volume}{33b}}, \bibinfo{pages}{606} (\bibinfo{year}{1978}).

\bibitem[{\citenamefont{Brock et~al.}(1994)\citenamefont{Brock, Greedan, and
  Kauzlarich}}]{SrMn2P2}
\bibinfo{author}{\bibfnamefont{S.~L.} \bibnamefont{Brock}},
  \bibinfo{author}{\bibfnamefont{J.~E.} \bibnamefont{Greedan}},
  \bibnamefont{and} \bibinfo{author}{\bibfnamefont{S.~M.}
  \bibnamefont{Kauzlarich}}, \bibinfo{journal}{Journal of Solid State
  Chemistry} \textbf{\bibinfo{volume}{113}}, \bibinfo{pages}{303 }
  (\bibinfo{year}{1994}).

\bibitem[{\citenamefont{Bobev et~al.}(2006)\citenamefont{Bobev, Merz, Lima,
  Fritsch, Thompson, Sarrao, Gillessen, and Dronskowski}}]{CaMn2Sb2}
\bibinfo{author}{\bibfnamefont{S.}~\bibnamefont{Bobev}},
  \bibinfo{author}{\bibfnamefont{J.}~\bibnamefont{Merz}},
  \bibinfo{author}{\bibfnamefont{A.}~\bibnamefont{Lima}},
  \bibinfo{author}{\bibfnamefont{V.}~\bibnamefont{Fritsch}},
  \bibinfo{author}{\bibfnamefont{J.~D.} \bibnamefont{Thompson}},
  \bibinfo{author}{\bibfnamefont{J.~L.} \bibnamefont{Sarrao}},
  \bibinfo{author}{\bibfnamefont{M.}~\bibnamefont{Gillessen}},
  \bibnamefont{and}
  \bibinfo{author}{\bibfnamefont{R.}~\bibnamefont{Dronskowski}},
  \bibinfo{journal}{Inorganic Chemistry} \textbf{\bibinfo{volume}{45}},
  \bibinfo{pages}{4047} (\bibinfo{year}{2006}).

\bibitem[{\citenamefont{Sangeetha et~al.}(2016)\citenamefont{Sangeetha, Pandey,
  Benson, and Johnston}}]{Sangeetha2016}
\bibinfo{author}{\bibfnamefont{N.~S.} \bibnamefont{Sangeetha}},
  \bibinfo{author}{\bibfnamefont{A.}~\bibnamefont{Pandey}},
  \bibinfo{author}{\bibfnamefont{Z.~A.} \bibnamefont{Benson}},
  \bibnamefont{and} \bibinfo{author}{\bibfnamefont{D.~C.}
  \bibnamefont{Johnston}}, \bibinfo{journal}{Phys. Rev. B}
  \textbf{\bibinfo{volume}{94}}, \bibinfo{pages}{094417}
  (\bibinfo{year}{2016}).

\bibitem[{\citenamefont{Fang et~al.}(2008)\citenamefont{Fang, Yao, Tsai, Hu,
  and Kivelson}}]{Fang2008}
\bibinfo{author}{\bibfnamefont{C.}~\bibnamefont{Fang}},
  \bibinfo{author}{\bibfnamefont{H.}~\bibnamefont{Yao}},
  \bibinfo{author}{\bibfnamefont{W.~F.} \bibnamefont{Tsai}},
  \bibinfo{author}{\bibfnamefont{J.~P.} \bibnamefont{Hu}}, \bibnamefont{and}
  \bibinfo{author}{\bibfnamefont{S.~A.} \bibnamefont{Kivelson}},
  \bibinfo{journal}{Physical Review B} \textbf{\bibinfo{volume}{77}}
  (\bibinfo{year}{2008}).

\bibitem[{\citenamefont{Fang et~al.}(2009)\citenamefont{Fang, Bernevig, and
  Hu}}]{Fang2009}
\bibinfo{author}{\bibfnamefont{C.}~\bibnamefont{Fang}},
  \bibinfo{author}{\bibfnamefont{B.~A.} \bibnamefont{Bernevig}},
  \bibnamefont{and} \bibinfo{author}{\bibfnamefont{J.~P.} \bibnamefont{Hu}},
  \bibinfo{journal}{Epl} \textbf{\bibinfo{volume}{86}} (\bibinfo{year}{2009}).

\bibitem[{\citenamefont{Johnston}(2010)}]{reviewDCJohnston}
\bibinfo{author}{\bibfnamefont{D.~C.} \bibnamefont{Johnston}},
  \bibinfo{journal}{Advances in Physics} \textbf{\bibinfo{volume}{59}},
  \bibinfo{pages}{803} (\bibinfo{year}{2010}).

\bibitem[{\citenamefont{Holt et~al.}(2011)\citenamefont{Holt, Sushkov, Stanek,
  and Uhrig}}]{Holt2011}
\bibinfo{author}{\bibfnamefont{M.}~\bibnamefont{Holt}},
  \bibinfo{author}{\bibfnamefont{O.~P.} \bibnamefont{Sushkov}},
  \bibinfo{author}{\bibfnamefont{D.}~\bibnamefont{Stanek}}, \bibnamefont{and}
  \bibinfo{author}{\bibfnamefont{G.~S.} \bibnamefont{Uhrig}},
  \bibinfo{journal}{Phys. Rev. B} \textbf{\bibinfo{volume}{83}},
  \bibinfo{pages}{144528} (\bibinfo{year}{2011}).

\bibitem[{\citenamefont{Stanek et~al.}(2011)\citenamefont{Stanek, Sushkov, and
  Uhrig}}]{Stanek2011}
\bibinfo{author}{\bibfnamefont{D.}~\bibnamefont{Stanek}},
  \bibinfo{author}{\bibfnamefont{O.~P.} \bibnamefont{Sushkov}},
  \bibnamefont{and} \bibinfo{author}{\bibfnamefont{G.~S.} \bibnamefont{Uhrig}},
  \bibinfo{journal}{Phys. Rev. B} \textbf{\bibinfo{volume}{84}},
  \bibinfo{pages}{064505} (\bibinfo{year}{2011}).

\bibitem[{\citenamefont{Johnston et~al.}(2011)\citenamefont{Johnston,
  McQueeney, Lake, Honecker, Zhitomirsky, Nath, Furukawa, Antropov, and
  Singh}}]{Johnston2011}
\bibinfo{author}{\bibfnamefont{D.~C.} \bibnamefont{Johnston}},
  \bibinfo{author}{\bibfnamefont{R.~J.} \bibnamefont{McQueeney}},
  \bibinfo{author}{\bibfnamefont{B.}~\bibnamefont{Lake}},
  \bibinfo{author}{\bibfnamefont{A.}~\bibnamefont{Honecker}},
  \bibinfo{author}{\bibfnamefont{M.~E.} \bibnamefont{Zhitomirsky}},
  \bibinfo{author}{\bibfnamefont{R.}~\bibnamefont{Nath}},
  \bibinfo{author}{\bibfnamefont{Y.}~\bibnamefont{Furukawa}},
  \bibinfo{author}{\bibfnamefont{V.~P.} \bibnamefont{Antropov}},
  \bibnamefont{and} \bibinfo{author}{\bibfnamefont{Y.}~\bibnamefont{Singh}},
  \bibinfo{journal}{Phys. Rev. B} \textbf{\bibinfo{volume}{84}},
  \bibinfo{pages}{094445} (\bibinfo{year}{2011}).

\bibitem[{\citenamefont{Singh et~al.}(2009{\natexlab{b}})\citenamefont{Singh,
  Sefat, McGuire, Sales, Mandrus, VanBebber, and Keppens}}]{Singh2009}
\bibinfo{author}{\bibfnamefont{D.~J.} \bibnamefont{Singh}},
  \bibinfo{author}{\bibfnamefont{A.~S.} \bibnamefont{Sefat}},
  \bibinfo{author}{\bibfnamefont{M.~A.} \bibnamefont{McGuire}},
  \bibinfo{author}{\bibfnamefont{B.~C.} \bibnamefont{Sales}},
  \bibinfo{author}{\bibfnamefont{D.}~\bibnamefont{Mandrus}},
  \bibinfo{author}{\bibfnamefont{L.~H.} \bibnamefont{VanBebber}},
  \bibnamefont{and} \bibinfo{author}{\bibfnamefont{V.}~\bibnamefont{Keppens}},
  \bibinfo{journal}{Phys. Rev. B} \textbf{\bibinfo{volume}{79}},
  \bibinfo{pages}{094429} (\bibinfo{year}{2009}{\natexlab{b}}).

\bibitem[{\citenamefont{Filsinger et~al.}(2017)\citenamefont{Filsinger,
  Schnelle, Adler, Fecher, Reehuis, Hoser, Hoffmann, Werner, Greenblatt, and
  Felser}}]{Filsinger2017}
\bibinfo{author}{\bibfnamefont{K.~A.} \bibnamefont{Filsinger}},
  \bibinfo{author}{\bibfnamefont{W.}~\bibnamefont{Schnelle}},
  \bibinfo{author}{\bibfnamefont{P.}~\bibnamefont{Adler}},
  \bibinfo{author}{\bibfnamefont{G.~H.} \bibnamefont{Fecher}},
  \bibinfo{author}{\bibfnamefont{M.}~\bibnamefont{Reehuis}},
  \bibinfo{author}{\bibfnamefont{A.}~\bibnamefont{Hoser}},
  \bibinfo{author}{\bibfnamefont{J.-U.} \bibnamefont{Hoffmann}},
  \bibinfo{author}{\bibfnamefont{P.}~\bibnamefont{Werner}},
  \bibinfo{author}{\bibfnamefont{M.}~\bibnamefont{Greenblatt}},
  \bibnamefont{and} \bibinfo{author}{\bibfnamefont{C.}~\bibnamefont{Felser}},
  \bibinfo{journal}{Phys. Rev. B} \textbf{\bibinfo{volume}{95}},
  \bibinfo{pages}{184414} (\bibinfo{year}{2017}).

\bibitem[{\citenamefont{An et~al.}(2009)\citenamefont{An, Sefat, Singh, and
  Du}}]{AnJiming2009}
\bibinfo{author}{\bibfnamefont{J.}~\bibnamefont{An}},
  \bibinfo{author}{\bibfnamefont{A.~S.} \bibnamefont{Sefat}},
  \bibinfo{author}{\bibfnamefont{D.~J.} \bibnamefont{Singh}}, \bibnamefont{and}
  \bibinfo{author}{\bibfnamefont{M.-H.} \bibnamefont{Du}},
  \bibinfo{journal}{Phys. Rev. B} \textbf{\bibinfo{volume}{79}},
  \bibinfo{pages}{075120} (\bibinfo{year}{2009}).

\bibitem[{\citenamefont{Ma et~al.}(2010)\citenamefont{Ma, Lu, and
  Xiang}}]{Ma2010}
\bibinfo{author}{\bibfnamefont{F.-j.} \bibnamefont{Ma}},
  \bibinfo{author}{\bibfnamefont{Z.-y.} \bibnamefont{Lu}}, \bibnamefont{and}
  \bibinfo{author}{\bibfnamefont{T.}~\bibnamefont{Xiang}},
  \bibinfo{journal}{Frontiers of Physics in China}
  \textbf{\bibinfo{volume}{5}}, \bibinfo{pages}{150} (\bibinfo{year}{2010}).

\bibitem[{\citenamefont{Xu et~al.}(2008)\citenamefont{Xu, Zhang, Dai, and
  Fang}}]{XuGang2008}
\bibinfo{author}{\bibfnamefont{G.}~\bibnamefont{Xu}},
  \bibinfo{author}{\bibfnamefont{H.}~\bibnamefont{Zhang}},
  \bibinfo{author}{\bibfnamefont{X.}~\bibnamefont{Dai}}, \bibnamefont{and}
  \bibinfo{author}{\bibfnamefont{Z.}~\bibnamefont{Fang}}, \bibinfo{journal}{EPL
  (Europhysics Letters)} \textbf{\bibinfo{volume}{84}}, \bibinfo{pages}{67015}
  (\bibinfo{year}{2008}).

\bibitem[{\citenamefont{Zbiri et~al.}(2009)\citenamefont{Zbiri, Schober,
  Johnson, Rols, Mittal, Su, Rotter, and Johrendt}}]{Zbiri2009}
\bibinfo{author}{\bibfnamefont{M.}~\bibnamefont{Zbiri}},
  \bibinfo{author}{\bibfnamefont{H.}~\bibnamefont{Schober}},
  \bibinfo{author}{\bibfnamefont{M.~R.} \bibnamefont{Johnson}},
  \bibinfo{author}{\bibfnamefont{S.}~\bibnamefont{Rols}},
  \bibinfo{author}{\bibfnamefont{R.}~\bibnamefont{Mittal}},
  \bibinfo{author}{\bibfnamefont{Y.}~\bibnamefont{Su}},
  \bibinfo{author}{\bibfnamefont{M.}~\bibnamefont{Rotter}}, \bibnamefont{and}
  \bibinfo{author}{\bibfnamefont{D.}~\bibnamefont{Johrendt}},
  \bibinfo{journal}{Phys. Rev. B} \textbf{\bibinfo{volume}{79}},
  \bibinfo{pages}{064511} (\bibinfo{year}{2009}).

\bibitem[{\citenamefont{Yin and Pickett}(2009)}]{Yin2009}
\bibinfo{author}{\bibfnamefont{Z.~P.} \bibnamefont{Yin}} \bibnamefont{and}
  \bibinfo{author}{\bibfnamefont{W.~E.} \bibnamefont{Pickett}},
  \bibinfo{journal}{Phys. Rev. B} \textbf{\bibinfo{volume}{80}},
  \bibinfo{pages}{144522} (\bibinfo{year}{2009}).

\bibitem[{\citenamefont{Boeri et~al.}(2010)\citenamefont{Boeri, Calandra,
  Mazin, Dolgov, and Mauri}}]{Boeri2010}
\bibinfo{author}{\bibfnamefont{L.}~\bibnamefont{Boeri}},
  \bibinfo{author}{\bibfnamefont{M.}~\bibnamefont{Calandra}},
  \bibinfo{author}{\bibfnamefont{I.~I.} \bibnamefont{Mazin}},
  \bibinfo{author}{\bibfnamefont{O.~V.} \bibnamefont{Dolgov}},
  \bibnamefont{and} \bibinfo{author}{\bibfnamefont{F.}~\bibnamefont{Mauri}},
  \bibinfo{journal}{Phys. Rev. B} \textbf{\bibinfo{volume}{82}},
  \bibinfo{pages}{020506} (\bibinfo{year}{2010}).

\bibitem[{\citenamefont{Han et~al.}(2009)\citenamefont{Han, Yin, Pickett, and
  Savrasov}}]{Han2009}
\bibinfo{author}{\bibfnamefont{M.~J.} \bibnamefont{Han}},
  \bibinfo{author}{\bibfnamefont{Q.}~\bibnamefont{Yin}},
  \bibinfo{author}{\bibfnamefont{W.~E.} \bibnamefont{Pickett}},
  \bibnamefont{and} \bibinfo{author}{\bibfnamefont{S.~Y.}
  \bibnamefont{Savrasov}}, \bibinfo{journal}{Phys. Rev. Lett.}
  \textbf{\bibinfo{volume}{102}}, \bibinfo{pages}{107003}
  (\bibinfo{year}{2009}).

\bibitem[{\citenamefont{Sefat et~al.}(2009{\natexlab{b}})\citenamefont{Sefat,
  Singh, Jin, McGuire, Sales, and Mandrus}}]{Sefat2009-Co}
\bibinfo{author}{\bibfnamefont{A.~S.} \bibnamefont{Sefat}},
  \bibinfo{author}{\bibfnamefont{D.~J.} \bibnamefont{Singh}},
  \bibinfo{author}{\bibfnamefont{R.}~\bibnamefont{Jin}},
  \bibinfo{author}{\bibfnamefont{M.~A.} \bibnamefont{McGuire}},
  \bibinfo{author}{\bibfnamefont{B.~C.} \bibnamefont{Sales}}, \bibnamefont{and}
  \bibinfo{author}{\bibfnamefont{D.}~\bibnamefont{Mandrus}},
  \bibinfo{journal}{Phys. Rev. B} \textbf{\bibinfo{volume}{79}},
  \bibinfo{pages}{024512} (\bibinfo{year}{2009}{\natexlab{b}}).

\bibitem[{\citenamefont{Zhou et~al.}(2011)\citenamefont{Zhou, Xu, Zhang, Xu,
  He, Yang, Chen, Xie, Cui, Arita et~al.}}]{NiARPESNoMoment}
\bibinfo{author}{\bibfnamefont{B.}~\bibnamefont{Zhou}},
  \bibinfo{author}{\bibfnamefont{M.}~\bibnamefont{Xu}},
  \bibinfo{author}{\bibfnamefont{Y.}~\bibnamefont{Zhang}},
  \bibinfo{author}{\bibfnamefont{G.}~\bibnamefont{Xu}},
  \bibinfo{author}{\bibfnamefont{C.}~\bibnamefont{He}},
  \bibinfo{author}{\bibfnamefont{L.~X.} \bibnamefont{Yang}},
  \bibinfo{author}{\bibfnamefont{F.}~\bibnamefont{Chen}},
  \bibinfo{author}{\bibfnamefont{B.~P.} \bibnamefont{Xie}},
  \bibinfo{author}{\bibfnamefont{X.-Y.} \bibnamefont{Cui}},
  \bibinfo{author}{\bibfnamefont{M.}~\bibnamefont{Arita}},
  \bibnamefont{et~al.}, \bibinfo{journal}{Phys. Rev. B}
  \textbf{\bibinfo{volume}{83}}, \bibinfo{pages}{035110}
  (\bibinfo{year}{2011}).

\bibitem[{\citenamefont{Richard et~al.}(2017)\citenamefont{Richard, van
  Roekeghem, Lv, Qian, Kim, Hoesch, Hu, Sefat, Biermann, and
  Ding}}]{Richard2017}
\bibinfo{author}{\bibfnamefont{P.}~\bibnamefont{Richard}},
  \bibinfo{author}{\bibfnamefont{A.}~\bibnamefont{van Roekeghem}},
  \bibinfo{author}{\bibfnamefont{B.~Q.} \bibnamefont{Lv}},
  \bibinfo{author}{\bibfnamefont{T.}~\bibnamefont{Qian}},
  \bibinfo{author}{\bibfnamefont{T.~K.} \bibnamefont{Kim}},
  \bibinfo{author}{\bibfnamefont{M.}~\bibnamefont{Hoesch}},
  \bibinfo{author}{\bibfnamefont{J.-P.} \bibnamefont{Hu}},
  \bibinfo{author}{\bibfnamefont{A.~S.} \bibnamefont{Sefat}},
  \bibinfo{author}{\bibfnamefont{S.}~\bibnamefont{Biermann}}, \bibnamefont{and}
  \bibinfo{author}{\bibfnamefont{H.}~\bibnamefont{Ding}},
  \bibinfo{journal}{Phys. Rev. B} \textbf{\bibinfo{volume}{95}},
  \bibinfo{pages}{184516} (\bibinfo{year}{2017}).

\bibitem[{\citenamefont{Singh et~al.}(2009{\natexlab{c}})\citenamefont{Singh,
  Green, Huang, Kreyssig, McQueeney, Johnston, and Goldman}}]{Singh2009-Mn}
\bibinfo{author}{\bibfnamefont{Y.}~\bibnamefont{Singh}},
  \bibinfo{author}{\bibfnamefont{M.~A.} \bibnamefont{Green}},
  \bibinfo{author}{\bibfnamefont{Q.}~\bibnamefont{Huang}},
  \bibinfo{author}{\bibfnamefont{A.}~\bibnamefont{Kreyssig}},
  \bibinfo{author}{\bibfnamefont{R.~J.} \bibnamefont{McQueeney}},
  \bibinfo{author}{\bibfnamefont{D.~C.} \bibnamefont{Johnston}},
  \bibnamefont{and} \bibinfo{author}{\bibfnamefont{A.~I.}
  \bibnamefont{Goldman}}, \bibinfo{journal}{Phys. Rev. B}
  \textbf{\bibinfo{volume}{80}}, \bibinfo{pages}{100403}
  (\bibinfo{year}{2009}{\natexlab{c}}).

\bibitem[{\citenamefont{Huang et~al.}(2008)\citenamefont{Huang, Qiu, Bao,
  Green, Lynn, Gasparovic, Wu, Wu, and Chen}}]{Huang2008}
\bibinfo{author}{\bibfnamefont{Q.}~\bibnamefont{Huang}},
  \bibinfo{author}{\bibfnamefont{Y.}~\bibnamefont{Qiu}},
  \bibinfo{author}{\bibfnamefont{W.}~\bibnamefont{Bao}},
  \bibinfo{author}{\bibfnamefont{M.~A.} \bibnamefont{Green}},
  \bibinfo{author}{\bibfnamefont{J.~W.} \bibnamefont{Lynn}},
  \bibinfo{author}{\bibfnamefont{Y.~C.} \bibnamefont{Gasparovic}},
  \bibinfo{author}{\bibfnamefont{T.}~\bibnamefont{Wu}},
  \bibinfo{author}{\bibfnamefont{G.}~\bibnamefont{Wu}}, \bibnamefont{and}
  \bibinfo{author}{\bibfnamefont{X.~H.} \bibnamefont{Chen}},
  \bibinfo{journal}{Phys. Rev. Lett.} \textbf{\bibinfo{volume}{101}},
  \bibinfo{pages}{257003} (\bibinfo{year}{2008}).

\bibitem[{\citenamefont{Hu et~al.}(2012)\citenamefont{Hu, Xu, Liu, Hao, and
  Wang}}]{Hu2012u}
\bibinfo{author}{\bibfnamefont{J.}~\bibnamefont{Hu}},
  \bibinfo{author}{\bibfnamefont{B.}~\bibnamefont{Xu}},
  \bibinfo{author}{\bibfnamefont{W.}~\bibnamefont{Liu}},
  \bibinfo{author}{\bibfnamefont{N.-N.} \bibnamefont{Hao}}, \bibnamefont{and}
  \bibinfo{author}{\bibfnamefont{Y.}~\bibnamefont{Wang}},
  \bibinfo{journal}{Phys. Rev. B} \textbf{\bibinfo{volume}{85}},
  \bibinfo{pages}{144403} (\bibinfo{year}{2012}).

\bibitem[{\citenamefont{McNally et~al.}(2015)\citenamefont{McNally, Simonson,
  Kistner-Morris, Smith, Hassinger, DeBeer-Schmitt, Kolesnikov, Zaliznyak, and
  Aronson}}]{McNally2015}
\bibinfo{author}{\bibfnamefont{D.~E.} \bibnamefont{McNally}},
  \bibinfo{author}{\bibfnamefont{J.~W.} \bibnamefont{Simonson}},
  \bibinfo{author}{\bibfnamefont{J.~J.} \bibnamefont{Kistner-Morris}},
  \bibinfo{author}{\bibfnamefont{G.~J.} \bibnamefont{Smith}},
  \bibinfo{author}{\bibfnamefont{J.~E.} \bibnamefont{Hassinger}},
  \bibinfo{author}{\bibfnamefont{L.}~\bibnamefont{DeBeer-Schmitt}},
  \bibinfo{author}{\bibfnamefont{A.~I.} \bibnamefont{Kolesnikov}},
  \bibinfo{author}{\bibfnamefont{I.~A.} \bibnamefont{Zaliznyak}},
  \bibnamefont{and} \bibinfo{author}{\bibfnamefont{M.~C.}
  \bibnamefont{Aronson}}, \bibinfo{journal}{Phys. Rev. B}
  \textbf{\bibinfo{volume}{91}}, \bibinfo{pages}{180407}
  (\bibinfo{year}{2015}).

\bibitem[{\citenamefont{Rastelli et~al.}(1979)\citenamefont{Rastelli, Tassi,
  and Reatto}}]{Rastelli1979}
\bibinfo{author}{\bibfnamefont{E.}~\bibnamefont{Rastelli}},
  \bibinfo{author}{\bibfnamefont{A.}~\bibnamefont{Tassi}}, \bibnamefont{and}
  \bibinfo{author}{\bibfnamefont{L.}~\bibnamefont{Reatto}},
  \bibinfo{journal}{Physica B+C} \textbf{\bibinfo{volume}{97}},
  \bibinfo{pages}{1 } (\bibinfo{year}{1979}).

\bibitem[{\citenamefont{Fouet et~al.}(2001)\citenamefont{Fouet, Sindzingre, and
  Lhuillier}}]{Fouet2001}
\bibinfo{author}{\bibfnamefont{J.~B.} \bibnamefont{Fouet}},
  \bibinfo{author}{\bibfnamefont{P.}~\bibnamefont{Sindzingre}},
  \bibnamefont{and}
  \bibinfo{author}{\bibfnamefont{C.}~\bibnamefont{Lhuillier}},
  \bibinfo{journal}{The European Physical Journal B - Condensed Matter and
  Complex Systems} \textbf{\bibinfo{volume}{20}}, \bibinfo{pages}{241}
  (\bibinfo{year}{2001}).

\bibitem[{\citenamefont{Oitmaa and Singh}(2011)}]{Oitmaa2011}
\bibinfo{author}{\bibfnamefont{J.}~\bibnamefont{Oitmaa}} \bibnamefont{and}
  \bibinfo{author}{\bibfnamefont{R.~R.~P.} \bibnamefont{Singh}},
  \bibinfo{journal}{Phys. Rev. B} \textbf{\bibinfo{volume}{84}},
  \bibinfo{pages}{094424} (\bibinfo{year}{2011}).

\bibitem[{\citenamefont{Torikachvili et~al.}(2008)\citenamefont{Torikachvili,
  Bud'ko, Ni, and Canfield}}]{pressure2008}
\bibinfo{author}{\bibfnamefont{M.~S.} \bibnamefont{Torikachvili}},
  \bibinfo{author}{\bibfnamefont{S.~L.} \bibnamefont{Bud'ko}},
  \bibinfo{author}{\bibfnamefont{N.}~\bibnamefont{Ni}}, \bibnamefont{and}
  \bibinfo{author}{\bibfnamefont{P.~C.} \bibnamefont{Canfield}},
  \bibinfo{journal}{Phys. Rev. Lett.} \textbf{\bibinfo{volume}{101}},
  \bibinfo{pages}{057006} (\bibinfo{year}{2008}).

\bibitem[{\citenamefont{Park et~al.}(2008)\citenamefont{Park, Park, Lee,
  Klimczuk, Bauer, Ronning, and Thompson}}]{CaFe2As2Pressure}
\bibinfo{author}{\bibfnamefont{T.}~\bibnamefont{Park}},
  \bibinfo{author}{\bibfnamefont{E.}~\bibnamefont{Park}},
  \bibinfo{author}{\bibfnamefont{H.}~\bibnamefont{Lee}},
  \bibinfo{author}{\bibfnamefont{T.}~\bibnamefont{Klimczuk}},
  \bibinfo{author}{\bibfnamefont{E.~D.} \bibnamefont{Bauer}},
  \bibinfo{author}{\bibfnamefont{F.}~\bibnamefont{Ronning}}, \bibnamefont{and}
  \bibinfo{author}{\bibfnamefont{J.~D.} \bibnamefont{Thompson}},
  \bibinfo{journal}{Journal of Physics: Condensed Matter}
  \textbf{\bibinfo{volume}{20}}, \bibinfo{pages}{322204}
  (\bibinfo{year}{2008}).

\bibitem[{\citenamefont{Alireza et~al.}(2009)\citenamefont{Alireza, Ko,
  Gillett, Petrone, Cole, Lonzarich, and Sebastian}}]{SrBaFe2AS2Pressure}
\bibinfo{author}{\bibfnamefont{P.~L.} \bibnamefont{Alireza}},
  \bibinfo{author}{\bibfnamefont{Y.~T.~C.} \bibnamefont{Ko}},
  \bibinfo{author}{\bibfnamefont{J.}~\bibnamefont{Gillett}},
  \bibinfo{author}{\bibfnamefont{C.~M.} \bibnamefont{Petrone}},
  \bibinfo{author}{\bibfnamefont{J.~M.} \bibnamefont{Cole}},
  \bibinfo{author}{\bibfnamefont{G.~G.} \bibnamefont{Lonzarich}},
  \bibnamefont{and} \bibinfo{author}{\bibfnamefont{S.~E.}
  \bibnamefont{Sebastian}}, \bibinfo{journal}{Journal of Physics: Condensed
  Matter} \textbf{\bibinfo{volume}{21}}, \bibinfo{pages}{012208}
  (\bibinfo{year}{2009}).

\bibitem[{\citenamefont{Marzari and Vanderbilt}(1997)}]{wannier1997}
\bibinfo{author}{\bibfnamefont{N.}~\bibnamefont{Marzari}} \bibnamefont{and}
  \bibinfo{author}{\bibfnamefont{D.}~\bibnamefont{Vanderbilt}},
  \bibinfo{journal}{Physical Review B} \textbf{\bibinfo{volume}{56}},
  \bibinfo{pages}{12847} (\bibinfo{year}{1997}), \bibinfo{note}{pRB}.

\bibitem[{\citenamefont{Souza et~al.}(2001)\citenamefont{Souza, Marzari, and
  Vanderbilt}}]{wannier2001}
\bibinfo{author}{\bibfnamefont{I.}~\bibnamefont{Souza}},
  \bibinfo{author}{\bibfnamefont{N.}~\bibnamefont{Marzari}}, \bibnamefont{and}
  \bibinfo{author}{\bibfnamefont{D.}~\bibnamefont{Vanderbilt}},
  \bibinfo{journal}{Physical Review B} \textbf{\bibinfo{volume}{65}},
  \bibinfo{pages}{035109} (\bibinfo{year}{2001}), \bibinfo{note}{pRB}.

\bibitem[{\citenamefont{Kuroki et~al.}(2008)\citenamefont{Kuroki, Onari, Arita,
  Usui, Tanaka, Kontani, and Aoki}}]{Kurohi2008}
\bibinfo{author}{\bibfnamefont{K.}~\bibnamefont{Kuroki}},
  \bibinfo{author}{\bibfnamefont{S.}~\bibnamefont{Onari}},
  \bibinfo{author}{\bibfnamefont{R.}~\bibnamefont{Arita}},
  \bibinfo{author}{\bibfnamefont{H.}~\bibnamefont{Usui}},
  \bibinfo{author}{\bibfnamefont{Y.}~\bibnamefont{Tanaka}},
  \bibinfo{author}{\bibfnamefont{H.}~\bibnamefont{Kontani}}, \bibnamefont{and}
  \bibinfo{author}{\bibfnamefont{H.}~\bibnamefont{Aoki}},
  \bibinfo{journal}{Phys. Rev. Lett.} \textbf{\bibinfo{volume}{101}},
  \bibinfo{pages}{087004} (\bibinfo{year}{2008}).

\bibitem[{\citenamefont{Pickett}(1989)}]{rmpPickett1989}
\bibinfo{author}{\bibfnamefont{W.~E.} \bibnamefont{Pickett}},
  \bibinfo{journal}{Rev. Mod. Phys.} \textbf{\bibinfo{volume}{61}},
  \bibinfo{pages}{433} (\bibinfo{year}{1989}).

\bibitem[{\citenamefont{Hu and Ding}(2012)}]{Hu_Ding}
\bibinfo{author}{\bibfnamefont{J.}~\bibnamefont{Hu}} \bibnamefont{and}
  \bibinfo{author}{\bibfnamefont{H.}~\bibnamefont{Ding}}, \bibinfo{journal}{Sci
  Rep} \textbf{\bibinfo{volume}{2}}, \bibinfo{pages}{381}
  (\bibinfo{year}{2012}).

\bibitem[{\citenamefont{Davis and Lee}(2013)}]{Davis29102013}
\bibinfo{author}{\bibfnamefont{J.~C.~S.} \bibnamefont{Davis}} \bibnamefont{and}
  \bibinfo{author}{\bibfnamefont{D.-H.} \bibnamefont{Lee}},
  \bibinfo{journal}{Proceedings of the National Academy of Sciences}
  \textbf{\bibinfo{volume}{110}}, \bibinfo{pages}{17623}
  (\bibinfo{year}{2013}).

\end{thebibliography}

\end{document}